\documentclass{aastex631}

\accepted{October 26, 2025}

\begin{document}

\title{Spatio-Temporal Evolution of the March 2022 ICME Revealed by Multi-Point Observations of Forbush Decreases}

\author[0009-0002-2446-7345]{Gaku Kinoshita}
\affiliation{Department of Earth and Planetary Science, Graduate School of Science, The University of Tokyo, Tokyo, Japan}
\author[0000-0003-0277-3253]{Beatriz Sanchez-Cano}
\affiliation{School of Physics and Astronomy, University of Leicester, Leicester, United Kingdom}
\author[0000-0001-7998-1240]{Yoshizumi Miyoshi}
\affiliation{Institute for Space-Earth Environmental Research, Nagoya University, Japan}
\affiliation{Kyung Hee University, Swon, Korea}
\author[0000-0003-2361-5510]{Laura Rodríguez-García}
\affiliation{European Space Agency (ESA), European Space Astronomy Centre (ESAC), Camino Bajo del Castillo s/n, 28692 Villanueva de la Cañada, Madrid, Spain}
\author[0000-0002-4489-8073]{Emilia Kilpua}
\affiliation{University of Helsinki, Helsinki, Finland}
\author[0000-0001-6807-8494]{Benoit Lavraud}
\affiliation{Laboratoire d’astrophysique de Bordeaux, CNRS, University of Bordeaux, Pessac, France}
\author[0009-0009-2857-2362]{Mathias Rojo}
\affiliation{Institut de Recherche en Astrophysique et Planétologie (IRAP), CNRS-UPS-CNES, Toulouse, France}
\author[0000-0002-5712-9396]{Marco Pinto}
\affiliation{Laboratório de Instrumentação e Física Experimental de Partículas: Lisbon, PT}
\author[0000-0002-4001-6352]{Yuki Harada}
\affiliation{Department of Geophysics, Graduate School of Science, Kyoto University, Kyoto, Japan}
\author[0000-0002-2759-7682]{Go Murakami}
\affiliation{Institute of Space and Astronautical Science (ISAS), Japan Aerospace Exploration Agency (JAXA), Japan}
\author[0000-0002-1354-3544]{Yoshifumi Saito}
\affiliation{Institute of Space and Astronautical Science (ISAS), Japan Aerospace Exploration Agency (JAXA), Japan}
\author[0000-0001-8851-9146]{Shoichiro Yokota}
\affiliation{Institute of Space and Astronautical Science, Osaka University, Japan}
\author[0000-0001-7894-8246]{Daniel Heyner}
\affiliation{Institut für Geophysik und extraterrestrische Physik, Technische Universität Braunschweig, Braunschweig, Germany}
\author[0000-0002-8435-7220]{David Fischer}
\affiliation{Space Research Institute, Austrian Academy of Sciences, Graz, Austria}
\author[0000-0001-8017-5676]{Nicolas Andre}
\affiliation{Institut de Recherche en Astrophysique et Planétologie (IRAP), CNRS, CNES, Université Toulouse III, Toulouse, France}
\affiliation{Institut Supérieur de l'Aéronautique et de l'Espace (ISAE‐SUPAERO), Université de Toulouse, Toulouse, France}
\author[0000-0001-5451-9367]{Kazuo Yoshioka}
\affiliation{Department of Complexity Science and Engineering, Graduate School of Frontier Science, The University of Tokyo, Japan}


%% argument for \collaboration is the collaboration identifier. authors are

\begin{abstract}

Interplanetary coronal mass ejections (ICMEs) cause ``Forbush Decreases (FDs)" effects, which are local decreases in background galactic cosmic rays (GCR). Even though FDs can be observed with simple particle instruments, their amplitude and shape provide physical profiles of passing ICMEs. However, in some cases, previous statistical studies of the heliocentric distance dependence of FD changes associated with ICME propagation have found no strong correlation. We need the criteria for evaluating the relationship between ICMEs' structure and FD, necessary for FD's statistical analysis. This study investigates the effect of evolutions and interactions of ICMEs on FDs’ profiles in the inner Solar System, using multipoint comparisons. We focus on multipoint ICME observations by Solar Orbiter, BepiColombo, and near-Earth spacecraft from March 10-16, 2022, when these spacecraft were ideally located for studying the radial and longitudinal evolutions of ICME and accompanying FDs. We compared GCR variations with the multiple in-situ data and ICME model, clarifying the correspondence between the evolution of each ICME structure in radial and azimuthal directions and the depth and gradients of the FD. The radial comparison revealed decreases in FD intensities and gradients associated with the expansion of the ICME. The longitudinal difference found in FD intensity indicates longitudinal variations of the ICME’s shielding effect. These results suggest that accurate multi-point FD comparisons require determining the relationship between the observer's position and the inner structure of the passing ICMEs. 

\end{abstract}



\section{Introduction} \label{sec:intro}
An Interplanetary Coronal Mass Ejection (ICME) is a structure composed of a magnetic flux rope and embedded plasma ejected from the Sun. Upon reaching the terrestrial magnetosphere, ICMEs are known to trigger geomagnetic storms, making them important targets in space weather research \citep[e.g.,][]{miyoshi2005ring, kilpua2017coronal}. In situ observations typically classify ICMEs into two primary regions: the Magnetic Ejecta (ME), which corresponds to the ejected magnetic flux rope, and the sheath region, which consists of the heated and compressed preceding solar wind field piled up at the leading edge of the ME. The ME is characterized by strong, smooth magnetic fields, low proton temperatures, densities, and plasma betas compared to the ambient solar wind. If ME's magnetic field shows a smooth rotation, signifying that it has retained the original flux rope structure and is crossed close enough at the center, the ME is called a Magnetic Cloud \citep[MC:][]{burlaga1981magnetic, kataoka2006flux}. The sheath often exhibits enhanced magnetic fluctuations, elevated proton temperatures, high plasma density, and increased solar wind speed \citep{cane2000coronal, kilpua2017coronal}.

ICMEs undergo complex changes as they expand during their interplanetary propagation and due to interactions with the ambient solar wind and other ICMEs. Consequently, ICMEs may lose their hydrodynamic coherence as they move away from the Sun \citep{owens2017coronal}. A comprehensive understanding of the geometric and magnetic changes from mesoscale to large scales as ICMEs propagate remains an open question \citep[e.g.][]{palmerio2024mesoscale, witasse2017interplanetary}. For instance, \cite{davies2021situ} analyzed an ICME event from April 2020 using multi-point in situ measurements from Solar Orbiter and near-Earth spacecraft and optical observations from STEREO-A \citep{kaiser2008stereo}. Their comparison with models revealed discrepancies in the attenuation of magnetic field strength. This suggests that a simple self-similar cylindrical expansion cannot fully explain ICME evolution and must consider complex interactions with the ambient solar wind. As such, modeling efforts incorporating multi-spacecraft in situ observations have been proposed to improve our understanding of ICME evolution \citep[e.g.][]{von2021radial}.

Forbush Decreases (FDs) are transient reduction in galactic cosmic ray (GCR) intensity as the ICME passes over an observer \citep{forbush1937effects}, and they are valuable indirect methods for detecting ICMEs. GCRs are highly energetic particles, primarily protons, that originate outside the solar system and are accelerated by supernova explosions, typically in the range of hundreds of MeV to several GeV \citep{simpson1983elemental}. GCRs are reduced locally in sheaths due to turbulence-driven diffusion, and in magnetic ejecta due to shielding by strong and closed magnetic fields \citep{wibberenz1998transient}. As a result, within ICMEs, GCRs typically decrease rapidly over several hours and recover gradually over several days \citep[e.g.][]{janvier2021two}. Although relatively simple particle detectors can detect FDs, they can offer valuable insight into passing ICMEs. The magnitude and profile of FDs reflect the associated ICME structure, and they play an essential role in expanding global monitoring networks. One notable feature of FDs is the variation in the gradient of one-step or two-step structures. The two-step structure has traditionally been attributed to the passage of both the sheath and the ME, with each step corresponding to the two regions of ICMEs. In contrast, one-step events suggest that only one of the two regions was encountered \citep{kunow2006situ}. However, \cite{jordan2011revisiting} demonstrated that this model does not comprehensively explain all observed FD steps, emphasizing the importance of considering interplanetary conditions for each event to develop a more robust classification scheme. 

% The magnitude and profile of FDs reflect information about the associated ICME structure. Since relatively simple particle detectors can detect FDs, they play an essential role in expanding global ICME monitoring networks.

Historically, previous FD studies on the Earth have utilized large datasets of ground-based neutron monitor data to investigate the relationship between FD substructures and ICME components. However, with the growing number of deep space missions, obtaining FD measurements at various heliocentric distances has become possible. \cite{witasse2017interplanetary} analyzed a unique ICME observed continuously by nine spacecraft from Venus to Voyager 2 (0.72–110 au) and compared FD profiles at Mars (1.41 au), comet 67P/Churyumov-Gerasimenko (3.13 au), and Saturn (9.94 au). They found a trend where FDs became deeper, steeper, and shorter closer to the Sun due to ICMEs' expansion and deceleration with distance. Similarly, \cite{winslow2018opening} examined a single ICME-related FD observed at Mercury (0.332 au), the Moon (1.0 au), and Mars (1.657 au), supporting the same radial dependence. Notably, a two-step FD was detected only at Mercury, suggesting that the number of FD steps may also depend on heliocentric distance. In contrast, statistical analyses of multiple FD events have not consistently revealed such radial trends. For example, \cite{davies2023characterizing} investigated 47 FD events observed by Mercury orbiter ``MErcury Surface, Space ENvironment, GEochemistry and Ranging \citep[MESSENGER:][]{solomon2007messenger}" during its orbital phase at Mercury (0.3-0.47 au), and \cite{belov2023helios} analyzed 765 events from the Helios mission (0.28-1 au). While both studies explored the expected attenuation of FD depth with increasing solar distance, they found only weak correlations. \cite{davies2023characterizing} also reported that 72\% of the observed events at 0.3–0.47 au exhibited two-step FDs. They attempted to compare the step number in the radial direction using previous studies on Earth (1 au). However, the step number ratios in the previous studies on Earth varied widely  (70\% in \citeauthor{cane1994cosmic} \citeyear{cane1994cosmic}, versus 16.2\% in \citeauthor{jordan2011revisiting} \citeyear{jordan2011revisiting}), and they were unable to reach a definitive conclusion. One of the key limitations of these studies is that each spacecraft (or the Earth) may have intersected different parts of the ICME with FDs, so the evolution of corresponding ICME-FD structures could not have been properly tracked. Additionally, MESSENGER, which has been widely used for multi-point studies on FDs and ICMEs observations in the inner solar system \citep[e.g.,][]{winslow2018opening, rodriguez2022evidence, davies2023characterizing}, has difficulties with solar wind observations due to the limited field of view of the particle instrument, which is caused by the sun shield \citep{andrews2007energetic}. Additionally, the magnetic field data acquired by MESSENGER inside Mercury's magnetosphere must be removed to isolate the solar wind component, making the interplanetary magnetic field dataset discontinuous \citep{winslow2018opening}. Reliable investigations of ICME events require both magnetic field and plasma observations to distinguish between typical solar wind field and plasma characteristics \citep{winslow2015interplanetary}, therefore, we need more detailed studies to understand the structure of ICMEs and their relationship with FDs in the inner heliosphere.

The previous studies listed above represent critical milestones, offering initial insights into the relationship between large-scale ICME evolution and associated FDs. In this study, we revisit detailed event studies of multi-point FD changes, considering their correspondence with passing ICME structures and evolutions, aiming to define accurate criteria for future analysis on statistical multi-point FD comparisons. We select the event that occurred in March 2022 and was observed by Solar Orbiter, BepiColombo, and near Earth spacecraft in an ideal geometric configuration to investigate the evolution of the ICME in the radial and azimuthal directions. We then compared the three point in situ data for FDs, magnetic fields, and solar winds, obtained at each location. Section \ref{sec:event} introduces a March 2022 ICME event observed by three spacecraft aligned radially and longitudinally in pairs. Section \ref{sec:method} provides an overview of the instruments aboard each spacecraft. Section \ref{sec:insitu} presents detailed analyses of ICME structures at each observation point. Section \ref{sec:dis} compares these observations across radial and longitudinal directions to trace ICME evolution and FD variation. Our summary and conclusions are presented in Section \ref{sec:sum}.

\section{The ICME event in 2022/3} \label{sec:event}

\begin{figure}[ht]
    \centering
    \includegraphics[width=0.95\linewidth]{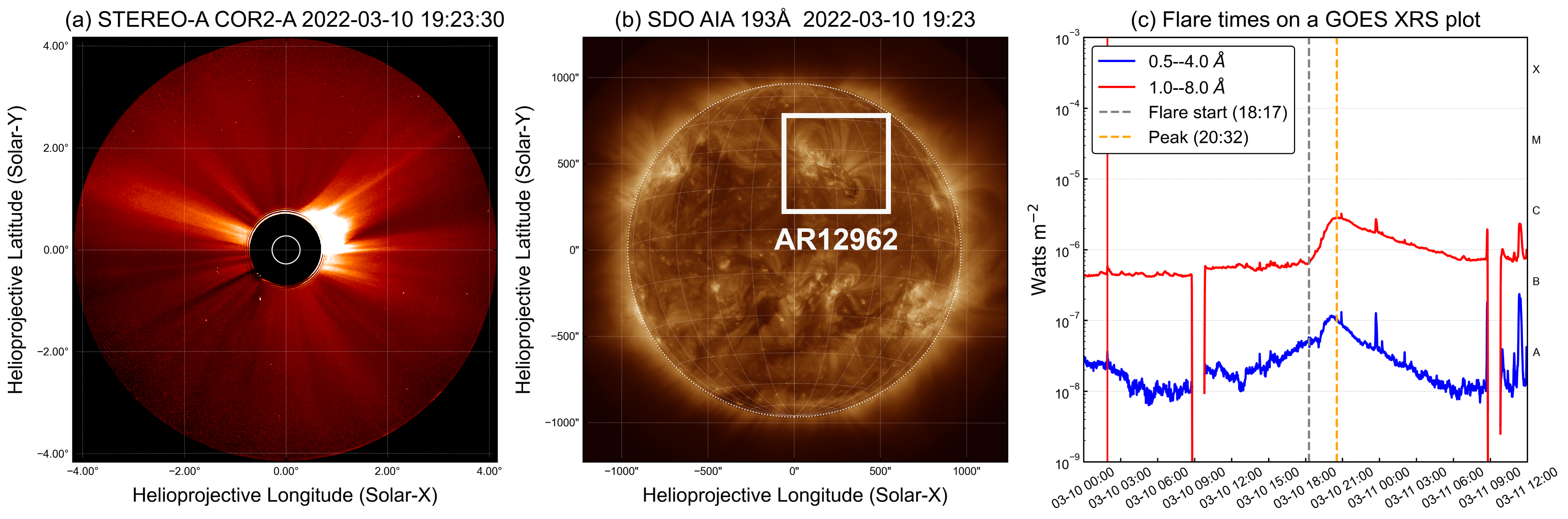}
    \caption{(a) STEREO-A SECCI/COR2 white-light image taken on 2022/3/10 at 19:23 UT (b) SDO AIA image taken on 2022/3/10 at 19:23 UT with 193 \AA, (c) GOES X-ray flux 1-minute data. These figures are created with SunPy software \citep{sunpy_community2020}}
    \label{fig:20220310_AR}
\end{figure}

\begin{figure}
    \centering
    \includegraphics[width=0.626\linewidth]{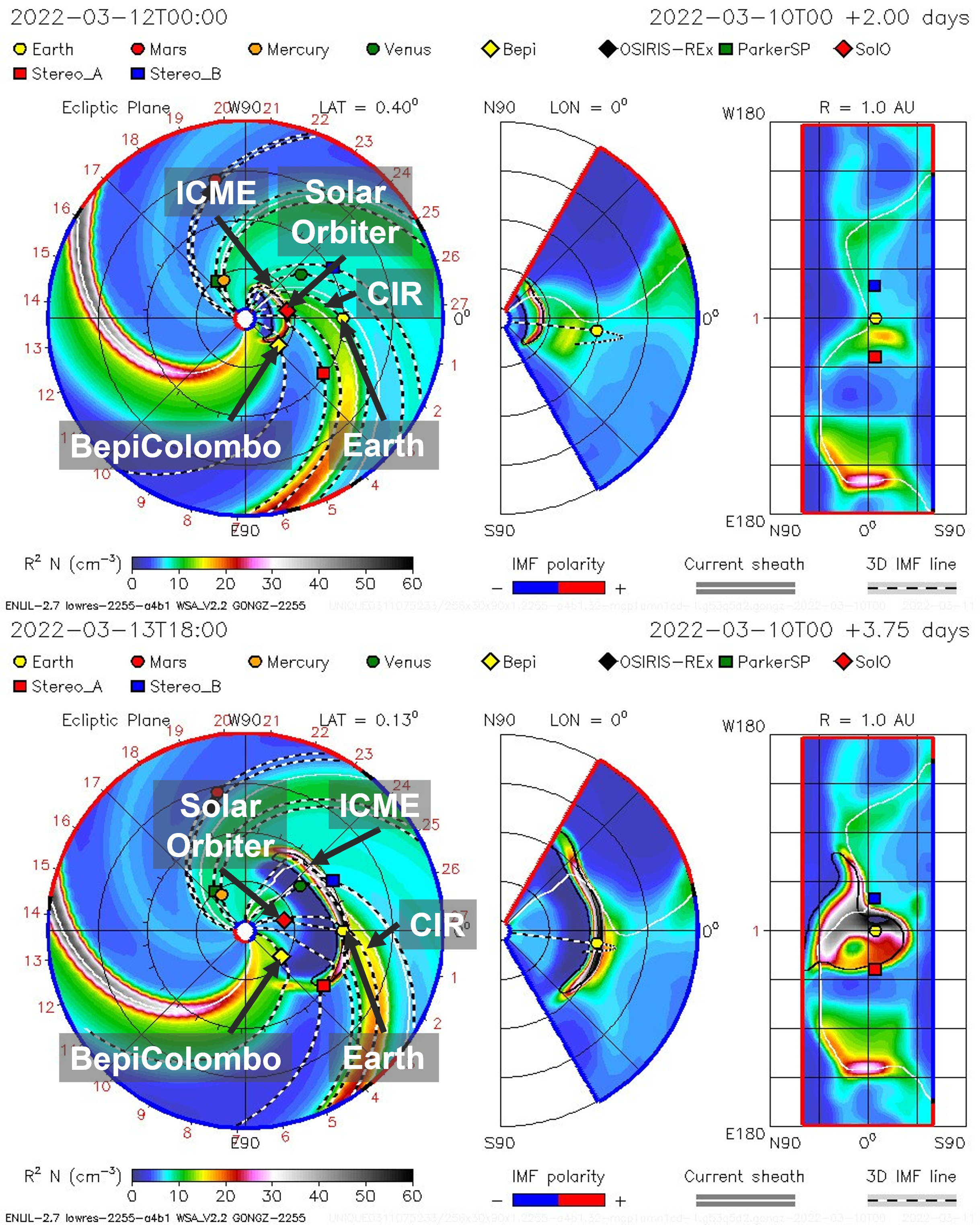}
    \caption{Simulations in the WSA-Enlil + Cone model for the ICME in 2022/3/12-3/13. The ICME simulations are adapted from the Space Weather Database Of Notifications, Knowledge, Information (DONKI). \url{https://kauai.ccmc.gsfc.nasa.gov/DONKI/view/WSA-ENLIL/19410/1}}
    \label{fig:20220312_cme}
\end{figure}

We mainly investigate an ICME that was observed in situ at multiple points in March 2022 and the accompanying FDs. This ICME was observed optically by STEREO-A SECCI/COR2 (Figure \ref{fig:20220310_AR}a) erupting from National Oceanic and Atmospheric Administration (NOAA) Active Region 12962 at 19:23 UT on March 10, 2022 (Figure \ref{fig:20220310_AR}b, \url{https://kauai.ccmc.gsfc.nasa.gov/DONKI/view/CME/19406/2}). The GOES satellite detected a related C2.8-class flare associated with this event starting at 18:17 UT, which peaked at 20:32 UT \citep[Figure \ref{fig:20220310_AR}c, \url{https://kauai.ccmc.gsfc.nasa.gov/DONKI/view/FLR/19413/3},][]{koya2024assessment}. 

To investigate the relationship between the propagating ICME and the positions of various spacecraft, we used the Wang-Sheeley-Arge (WSA)-ENLIL + Cone model, provided by the Community Coordinated Modeling Center (CCMC). This model uses the semi-empirical WSA model \citep[]{arge2000improvement, arge2004stream} to calculate the solar wind plasma and magnetic field conditions up to 21.5 solar radii, which serve as the inner boundary conditions for the three-dimensional Magnetohydrodynamics model ENLIL \citep{odstrcil2003modeling}.
The propagation of the ICME is then simulated by inserting a simplified Cone model \citep[]{xie2004cone, millward2013operational}. The ICME studied here passed through the locations of Solar Orbiter, Earth, and BepiColombo, as shown in Figure \ref{fig:20220312_cme}. 

Furthermore, Figure \ref{fig:20220312_cme} and spacecraft coordinates summarized in Table \ref{tab:location} show that Solar Orbiter and Earth were nearly aligned in a radial direction at different heliocentric distances, being there able to measure radial variations in the ICME's structure. In contrast, the Solar Orbiter and BepiColombo were at nearly the same heliocentric distance but separated by approximately 49 degrees in longitude, enabling the investigation of longitudinal differences in the ICME’s structure. 

Due to this ideal spacecraft configuration, this ICME has been analyzed in several previous studies.
\cite{jackson2023forecasting} reconstructed the ICME’s three-dimensional structure using interplanetary scintillation (IPS) observations, which utilize variations in received radio frequencies caused by plasma fluctuations along the line of sight between radio sources and ground stations. \cite{laker2024using} used Solar Orbiter data and modeling to study ICME arrival predictions. \cite{zhuang2024combining} examined the correspondence between in situ measurements from Solar Orbiter and optical observations from STEREO-A. \cite{beatriz2025cruise} focuses on BepiColombo's data from this event. In this study, we conduct a multi-point comparison of FDs of this event for the first time.

\begin{table}[ht]
 \caption{The locations of spacecraft in 10th Mar. 2022 calculated with Solar-Mach tool \citep{gieseler2023solar}}
 \centering
 \begin{tabular}{l c c c}
 \hline
    & Earth & Solar Orbiter &  BepiColombo \\
 \hline
   Carrington longitude [$^\circ$] & 300.6 &  310.7 & 262.0 \\
   Carrington latitude [$^\circ$]
  & -7.2 & -4.3 & 0.5 \\
       Heliocent. distance [au]
  & 0.99 & 0.44 & 0.43 \\
 \hline
 \label{tab:location}
 \end{tabular}
 \end{table}

\section{Data Sources} \label{sec:method}
\subsection{Solar orbiter} \label{sec:solo_method}
Solar Orbiter \citep{muller2020solar} is a solar observation spacecraft jointly operated by the European Space Agency (ESA) and the National Aeronautics and Space Administration (NASA).

The Energetic Particle Detector \citep[EPD:][]{rodriguez2020energetic} is a particle observation instrument mounted on Solar Orbiter. Among its components, the High Energy Telescope (HET) is responsible for observing high-energy particles such as Solar Energetic Particles (SEPs) and Galactic Cosmic Rays (GCRs). The HET is equipped with four semiconductor detectors (A1, A2, B1, B2) with a thickness of 300 $\mu$m and photodiodes (C1, C2) located on either side of 2 cm $\mathrm{Bi_4Ge_3O_{12}}$ (BGO) scintillators \citep[see Figure 31 of][]{rodriguez2020energetic}. The photodiodes convert scintillation light into electrical signals when particles strike the scintillators. Depending on the energy loss thresholds, there are high-gain and low-gain channels. The lower limits of detection of primary energies for each channel are 12 MeV for high-gain and 16 MeV for low-gain. C1 and C2 detectors can also be used as simple particle counters to achieve high count statistics, sacrificing information about energy and incident direction. Since GCRs arrive from all directions, and the C detectors are positioned at the center of an aluminum box, providing isotropic shielding, summing the counts from C1 and C2 for each channel can be considered a measure of GCR counts \citep{von2021radial}. This study uses low-gain counts with energy ranges similar to those of other particle detectors. The Solar Wind Analyzer \citep[SWA:][]{owen2020solar} is designed for in situ observation of solar wind plasma, comprising several types of sensors. In this study, we use the Proton-Alpha Sensor (SWA-PAS) and the Electron Analyzer System (SWA-EAS). The SWA-PAS is used for observations of solar wind protons and alpha particles. It measures them at a 4 s cadence with a maximum resolution of 96 energy steps, from 200 eV/e to 20 keV/e. The SWA-EAS measures solar wind electrons in the 1 eV to 5 keV range. It consists of two top-hat-type electrostatic analyzers, EAS1 and EAS2, which complement each other's fields of view. In this study, data from both analyzers were examined, but no significant differences in trends were observed; only the data from EAS1 are presented. The Radio and Plasma Wave instrument \citep[RPW:][]{maksimovic2020solar} can measure electron densities from probe-to-spacecraft potential \citep{khotyaintsev2021density}. MAG \citep{horbury2020solar} is a conventional dual fluxgate instrument that measures interplanetary magnetic fields.

\subsection{Near Earth spacecraft} \label{sec:earth_method}
For GCR data near the Earth, this study utilizes observations from the Cosmic Ray Telescope for the Effects of Radiation \citep[CRaTER;] []{spence2010crater} onboard NASA’s Lunar Reconnaissance Orbiter \citep[LRO;] []{vondrak2010lunar}. Since LRO operates in lunar orbit, its GCR measurements are less affected by the terrestrial magnetosphere and atmosphere comparing with ground observations on the Earth; therefore, CRaTER is suitable for direct multi-point FD comparisons. CRaTER is a radiation detector onboard the Lunar Reconnaissance Orbiter (LRO). It consists of three pairs of silicon detectors with thicknesses of 140 $\mu$m and 1000 $\mu$m, with tissue-equivalent plastic (TEP) absorbers placed between them \citep[see Figure 1 in ][]{wilson2020precise}.
This study utilizes the data named ``Good events" from CRaTER, which represents the total number of particles detected by any of the detectors. Since the threshold energy of the incident protons for any of the detectors is greater than 12.7 MeV, this data represents measurements of protons with an energy greater than 12.7 MeV. It exhibits good counting statistics, but it also has the disadvantage of a higher contamination by secondary particles from the lunar surface \citep{von2021radial}. 

LRO does not have plasma and magnetic field instruments, so this study uses magnetic field data from the Magnetic Field Investigation \citep[MFI;][]{lepping1995wind} and solar wind data from the Solar Wind Experiment \citep[SWE;][]{ogilvie1995swe} onboard Wind spacecraft \citep{ogilvie1997wind}, which is at the Lagrange point (L1). The distance between LRO, Wind, and the Earth is small enough compared to the typical scale of ICMEs, allowing the data from these spacecraft to be approximated as representative of the terrestrial space environment. Linking cosmic ray data from LRO or ground-based neutron monitors with plasma and magnetic field observations from other satellites near the Earth has been employed in previous studies \citep[e.g.,][]{janvier2021two, winslow2018opening}.

\begin{figure}[h]
    \centering
    \includegraphics[width=0.8\linewidth]{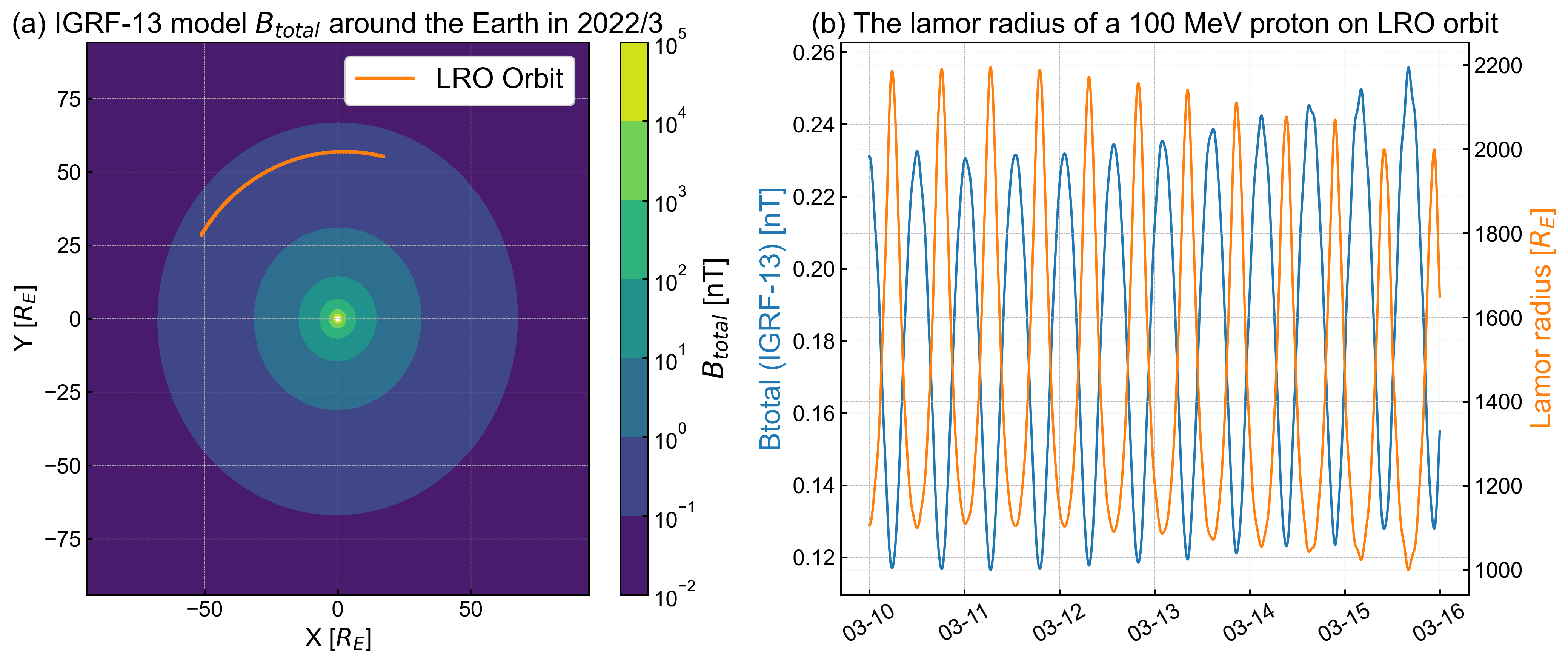}
    \caption{(a) IGRF-13 magnetic field distribution model and the orbit of LRO in the X-Y plane of J2000 coordinate system, (b) The $B_{total}$ variations of LRO's orbit (IGRF-13) and changes of lamor radius of a 100 MeV proton. $R_E$ represents the Earth's radius.}
    \label{fig:IGRF}
\end{figure}

Since the lunar orbit occasionally crosses into the terrestrial magnetosphere, an assessment of the influence of the terrestrial magnetic field on LRO’s orbit is necessary. Figure \ref{fig:IGRF}a presents the relationship between LRO’s trajectory and the magnetic field distribution from the International Geomagnetic Reference Field (IGRF-13) model \citep{ce1997international}, as well as a time series of the Larmor radii for 100 MeV protons calculated from the model magnetic field strength along the LRO orbit (Figure \ref{fig:IGRF}b). During the period surrounding this ICME event, the terrestrial magnetic field at LRO’s location remains on the order of $10^{-1}$ nT, and the Larmor radii of 100 MeV protons reach approximately thousands of Earth radii ($R_E$). These values indicate that the terrestrial magnetosphere is unlikely to significantly influence the behavior of GCRs in the energy range of several hundred MeV to several GeV. Therefore, comparing LRO’s GCRs data directly with the other interplanetary spacecraft is reasonable.

\begin{figure}
    \centering
    \includegraphics[width=0.75\linewidth]{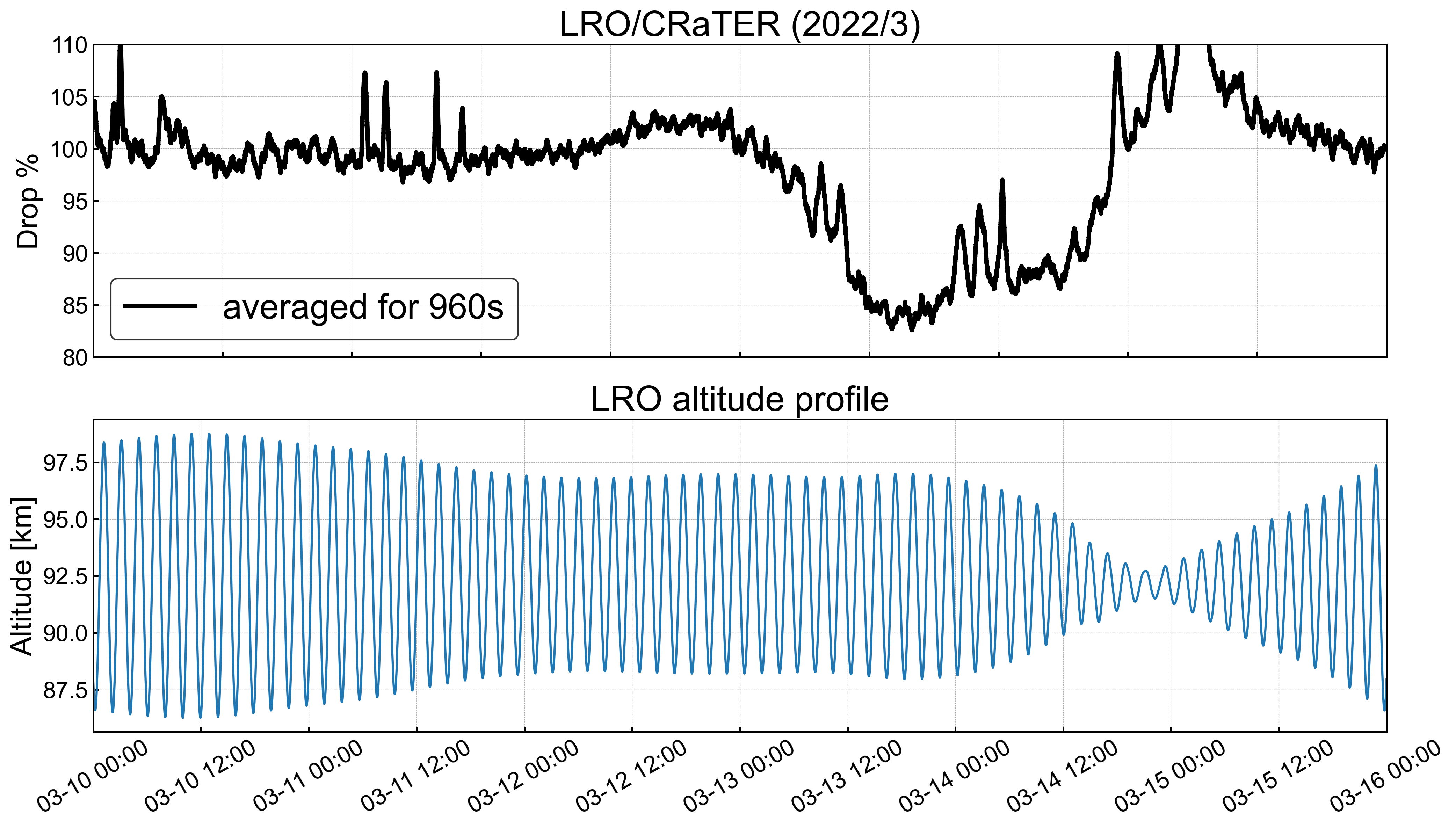}
    \caption{FD detected by CRaTER in March 2022 and lunar altitude variations of LRO}
    \label{fig:LRO_FD}
\end{figure}

Since LRO orbits the Moon in an elliptical trajectory, the fraction of CRaTER’s field of view occupied by the lunar surface varies with altitude. This leads to periodic cosmic ray count rate variations due to shielding and lunar surface reflection effects. These variations are unrelated to ICMEs, and previous studies on LRO’s FD events have applied different correction methods to remove them. For instance, \cite{von2021radial} used an empirical formula, while \cite{winslow2018opening} applied a Fourier-based low-pass filter. In contrast, for the FD event analyzed in this study (Figure \ref{fig:LRO_FD}), LRO’s altitude variations remain within 10 km, resulting in minimal high-frequency components in the data (For reference, during the event studied by \cite{von2021radial}, the altitude of LRO varied between 54 and 132 km.). As a result, the FD structure could be sufficiently extracted using a simple moving average, and no additional corrections were applied.

We use cosmic ray data from a ground-based neutron monitor to verify whether the FD observed by LRO was also detected on Earth's surface. To select the most suitable terrestrial data for comparison with spacecraft data, we referred to the specifications of various neutron monitors listed in Table 1 of \cite{jordan2011revisiting}. We chose data from the South Pole Neutron Monitor (SoPo), which has the lowest rigidity. Due to its proximity to the magnetic pole, atmospheric shielding is the dominant effect, resulting in a cutoff rigidity of 0.1 GV and a proton cutoff energy of 450 MeV \citep{clem2000neutron}.

To confirm the ICME's arrival at the  Earth, we utilize the Dst index and SYM/H index, both indicators of geomagnetic storm intensity \citep{wanliss2006high}.

\subsection{BepiColombo}
BepiColombo is a Mercury exploration mission jointly conducted by the Japan Aerospace Exploration Agency (JAXA) and ESA. The mission consists of two Mercury orbiters: the Mercury Magnetospheric Orbiter \citep[MMO:][]{go2020mio}, responsible for investigating Mercury’s magnetosphere, and the Mercury Planetary Orbiter (MPO) \citep{benkhoff2021bepicolombo}, dedicated to surface observations. Additionally, during the cruise phase only, the MMO is shielded from the solar radiation by the MMO Sunshield and Interface Structure (MOSIF), and the entire spacecraft is propelled by the Mercury Transfer Module (MTM) during the cruise phase to Mercury. Therefore, due to this stuck configuration during cruising, as will be explained in detail later, caution is required when interpreting the data because the field of view of the particle detector is limited. After the planned orbital insertion in 2026, the MMO and the MPO will be separated and conduct independent observations of Mercury.

\begin{figure}[h]
    \centering
    \includegraphics[width=0.6\linewidth]{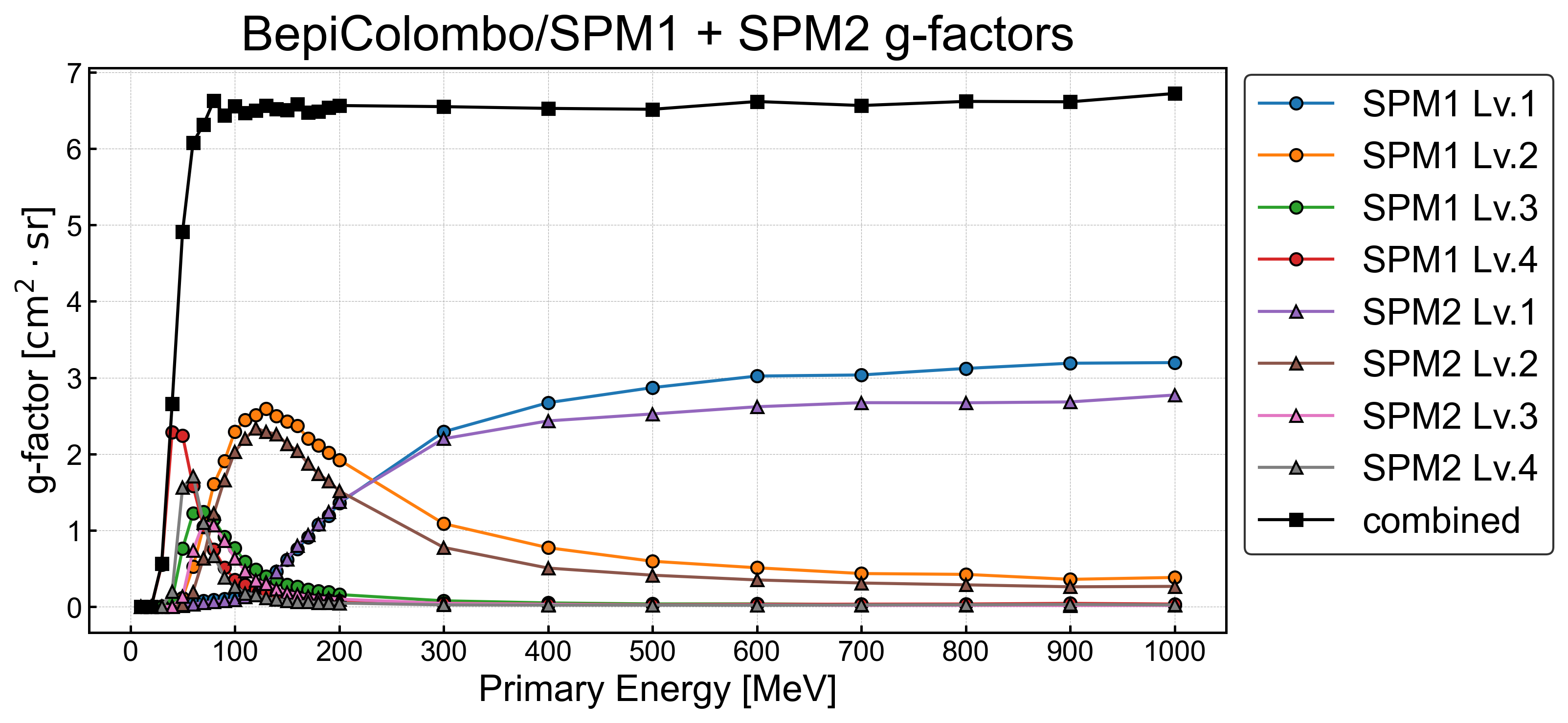}
    \caption{The response functions of MMO/SPM. $10^6$ protons are tested for each primary energy to derive g-factors. These response functions are derived based on the shielding effects around the MMO by the MOSIF, the MPO, and the MTM, so please note that they are only valid for events detected during the cruise phase.}
    \label{fig:SPM-function}
\end{figure}

The Solar Particle Monitor (SPM) \citep{kinoshita2025simulation}, onboard the MMO, is a radiation housekeeping instrument. It has two 10 mm $\times$ 10 mm $\times$ 0.3 mm Solid State Detectors (SSD), called SPM1 and SPM2. Each SSD measures the time series of the 32-second integrated count, classifying it into four energy channels (Lv.1 to Lv.4) based on deposited energy. Figure \ref{fig:SPM-function} shows the simulated response functions of SPM's physical model constructed by ``Geant4 \citep{allison2016recent}" radiation simulation toolkit (see also the description of Figure 3 in \cite{kinoshita2025simulation}). To achieve higher counting statistics, the sum of all channels from SPM1 and SPM2, represented by the black ``combined" function in Figure \ref{fig:SPM-function}, is used for multi-point comparisons. The vertical axis shows the geometrical factor \citep[g-factor:][]{sullivan1971geometric}, which is nearly equivalent to detection efficiency variations since the g-factors are derived by multiplying the detection efficiency by constants \citep[see Equation 2 in][]{kinoshita2025simulation}. This combined response function enables recording of integral counts of protons above 20 MeV. Figure 3 in \cite{von2021radial} shows the response functions for Solar Orbiter/HET and LRO/CRaTER. The g-factor is plotted on the vertical axis and primary energy on the horizontal axis, as in Figure \ref{fig:SPM-function} of this paper. As BepiColombo/SPM, the shape of each function is open, meaning that the g-factor does not decrease at high primary energies. As discussed in Section \ref{sec:solo_method} and \ref{sec:earth_method}, the lower limits of the response functions for Solar Orbiter/HET and LRO/CRaTER are 16 and 12.7 MeV, respectively. Since this paper uses Solar Orbiter/HET and LRO/CRaTER (lower limit: 16 MeV vs. 12.7 MeV) for radial comparison, and BepiColombo/SPM and Solar Orbiter/HET (20 MeV vs. 16 MeV) for azimuthal comparison, the differences in the measured energy ranges are minor, and a multi-point direct comparison is possible. Since this study focuses on the relative variations of cosmic ray background, the absolute differences in the g-factors are not significant issues.

For solar wind observations, this study utilizes the Mercury Ion Analyzer (MIA) and the Mercury Electron Analyzer (MEA), both part of the Mercury Plasma Particle Experiment \citep[MPPE:][]{saito2021pre} onboard the MMO. The MIA is a top-hat electrostatic analyzer that observes low-energy ions. One of the MIA data products is generated by summing measurements from four spin sectors (d1–4)\citep{fraenz2024spacecraft}, which are designed to cover different viewing directions but look at the same direction in the cruise phase configuration. We added the counts of d1 and d3 to achieve higher counting statistics, since d1 and d3 have the same sensitivity and energy steps. The MEA also uses a top-hat configuration with an energy resolution of 10\%. The field of view of MEA1 is 8° × 180° because the MAST-MGF \citep{baumjohann2020bepicolombo} shadows half of the detector \citep{rojo2024electron}. In this study, we used omnidirectional energy flux spectra with 16 energy steps, a duration of 4 s, spanning the energy range from 3 to 300 eV, and electronic densities. A crucial point when interpreting particle data onboard MMO during the cruise phase is that the field of view (FOV) is constrained due to the stuck configuration with MPO, MOSIF, and MTM as described above and in \cite{rojo2024electron, harada2024deep, harada2022bepicolombo, go2020mio}. Most of the particle instruments primarily observe the +Z direction in the spacecraft coordinate system because MOSIF is opened in the +Z direction and MPO \& MTM are combined in the -Z direction \citep[e.g.,][]{go2020mio, kinoshita2025simulation}. However, in interplanetary space, GCR incidence can be assumed to be isotropic. This limitation does not pose a significant issue for SPM's data since FD studies focus on relative variations rather than absolute values. The field of view limitation is more crucial for MPPE's low-energy solar wind observation. However, electrons pass through the ambient structure more easily than ions, reducing the effect of anisotropy. Therefore, depending on the direction of particle incidence, even if MEA observes a CME, there is no guarantee that MIA will detect it in the same way, and MIA may only record a weak enhancement or fail to detect the event completely. Concerning electron density, due to the limitations of the field of view mentioned above and the lack of actual measurements of the floating spacecraft potential during cruise, estimates were made using the procedure described in \cite{rojo2024electron}.

We use data from the fluxgate magnetometer MPO-MAG \citep{heyner2021bepicolombo}, which measures Mercury’s intrinsic and interplanetary magnetic fields.

\section{In Situ Observations} \label{sec:insitu}
\begin{figure}[htbp]
    \centering
    \includegraphics[width=0.85\linewidth]{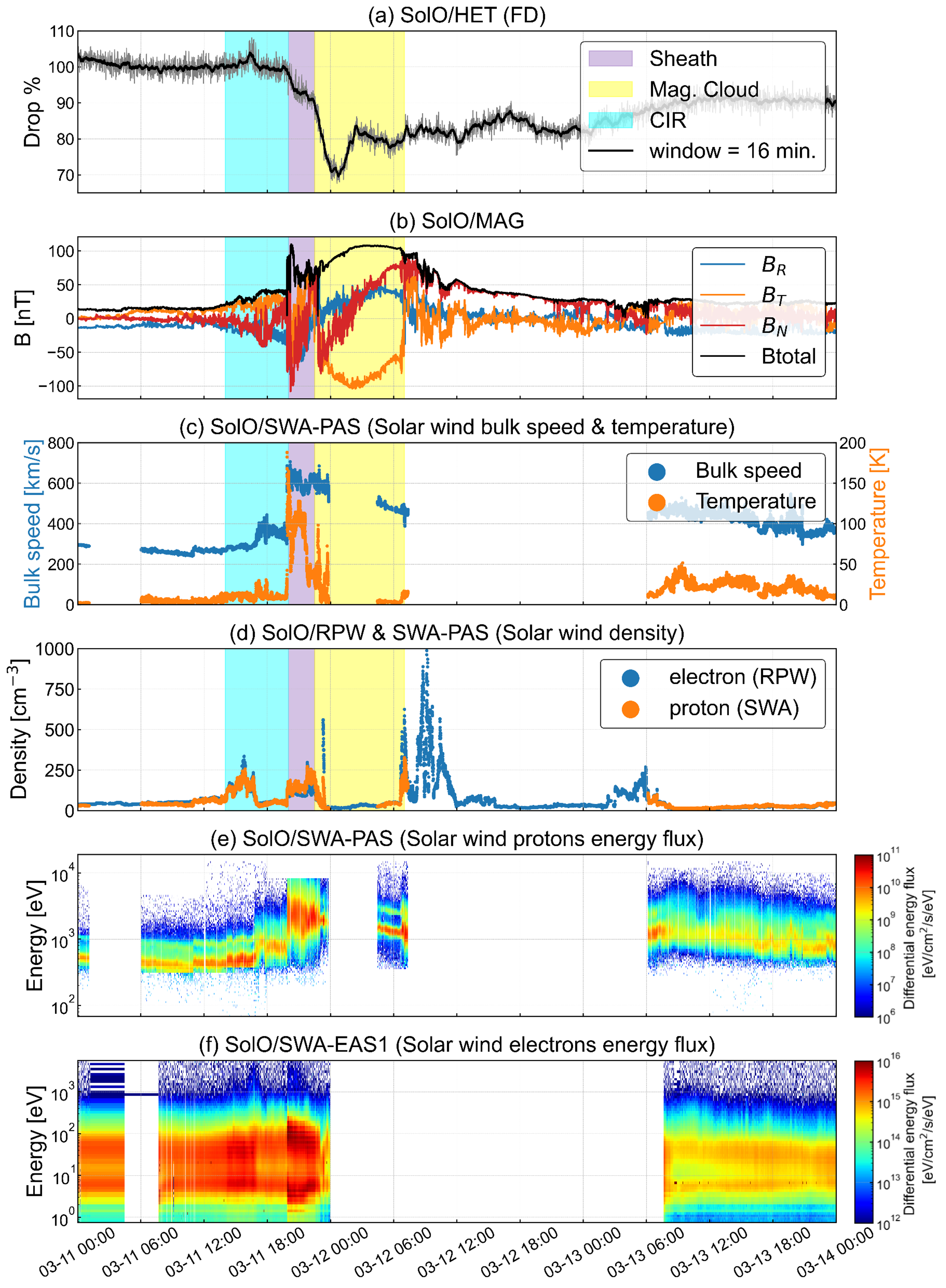}
    \caption{ICME observation data from Solar Orbiter in March 2022. (a): GCR variations (FD) normalized with the background before ICME arrival set to 100\% (HET-EPD). (b): magnetic field (RTN coordinate system), (c): solar wind velocity and temperature (SWA-PAS), (d): solar wind density (electron: RPW, proton: SWA-PAS), (e): solar wind proton energy flux (SWA-PAS), (f): solar wind electron energy flux (SWA-EAS1)}
    \label{fig:SolO}
\end{figure}\begin{figure}[h]
    \centering
    \includegraphics[width=0.8\linewidth]{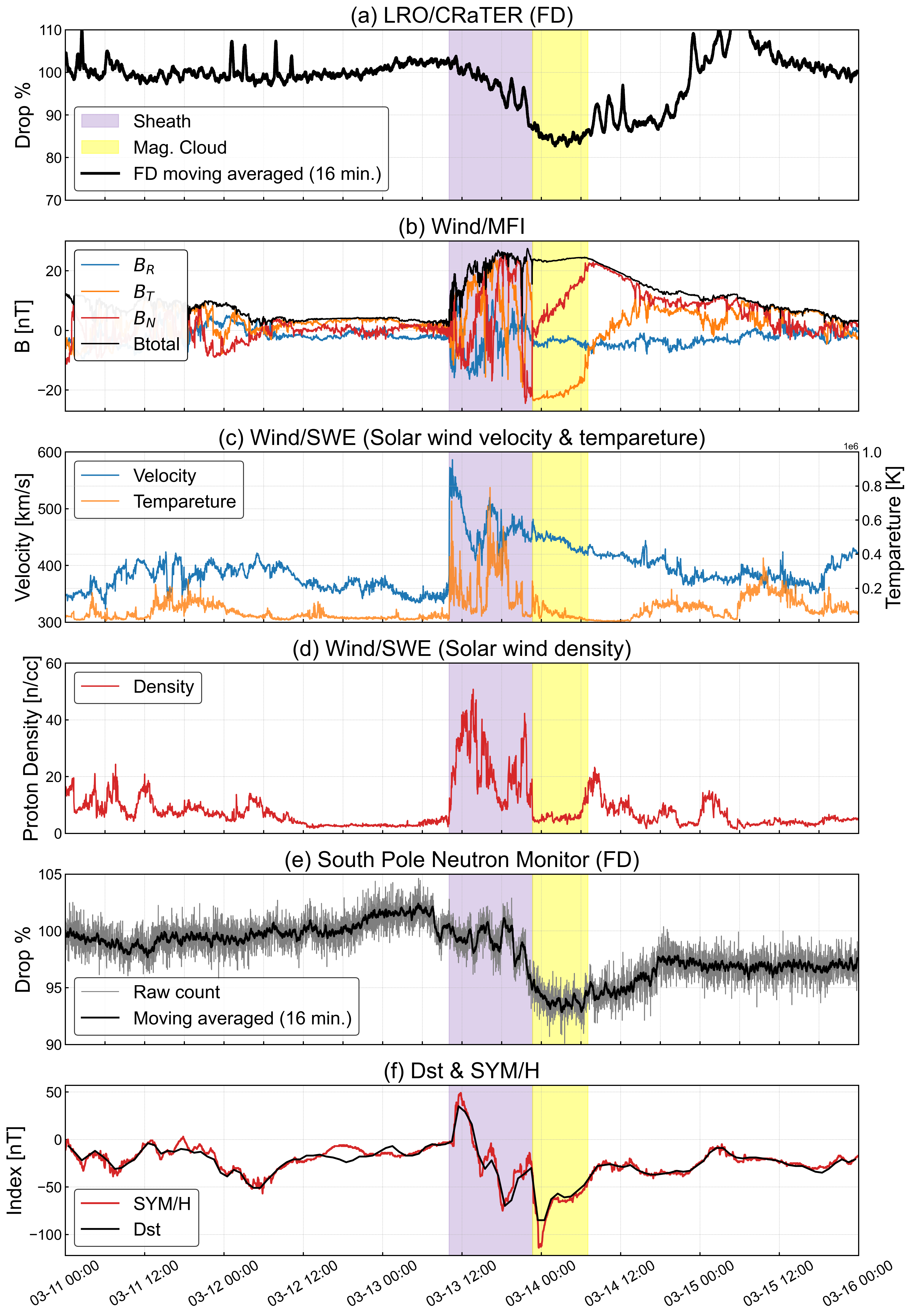}
    \caption{ICME observation data near Earth in 2022/3. (a): LRO/CRaTER's GCR variations (FD) normalized with the background before ICME arrival set to 100\% (the interplanetary observation), (b): Wind's magnetic field (RTN coordinate system), (c): Wind's solar wind speed and temperature, (d): Wind's solar wind density, (e): South Pole Neutron Monitor's GCR variations (FD) normalized with the background before ICME arrival set to 100\% (the ground observation), (f): The geomagnetic indices (Dst index and SYM/H index).}
    \label{fig:Earth}
\end{figure}

\begin{figure}[h]
    \centering
    \includegraphics[width=0.8\linewidth]{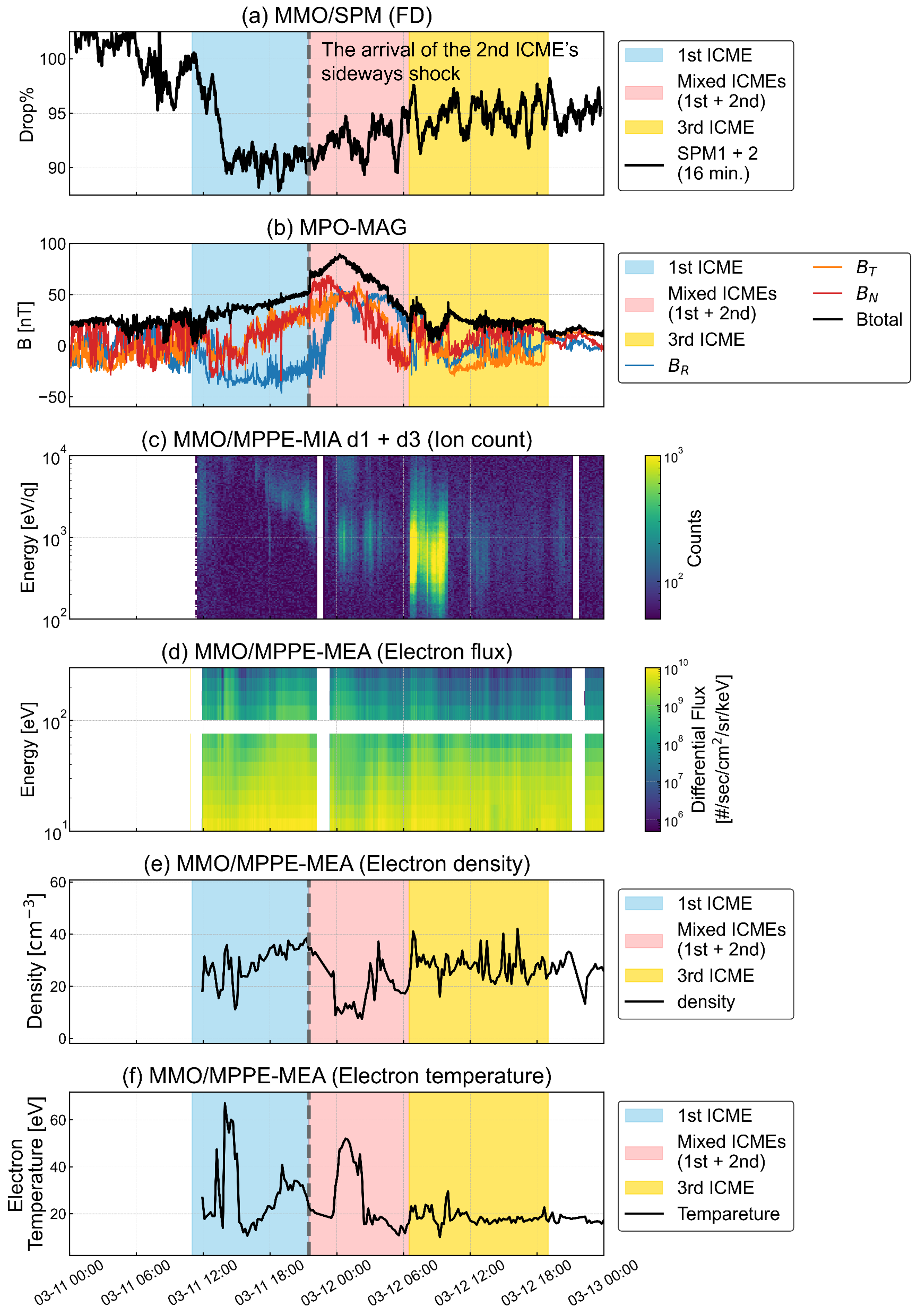}
    \caption{ICME observations of BepiColombo in 2022/3. (a): MMO/SPM's GCR variations (FD) normalized with the background before ICME arrival set to 100\%. To align with data from other probes, we use a 960 s (16 min.) moving average, when a moving average with a window = 30 was applied to the 32-second integrated SPM data, (b): MPO-MAG magnetic field (RTN coordinate system), (c): MMO/MIA's ion counts, (d): MMO/MEA's electron differential flux, (e): MMO/MEA's electron density, (f): MMO/MEA's electron temperature}
    \label{fig:Bepi-CME}
\end{figure}

\subsection{Solar Orbiter}\label{sec:solo}
Figure \ref{fig:SolO} presents observations from Solar Orbiter, which was located 0.44 au from the Sun. The ICME arrived at Solar Orbiter at 19:52 UT on March 11, 2022. The sheath (indicated in purple in Figure \ref{fig:SolO}) exhibits strong fluctuations in each magnetic field component and relatively high values of the total magnetic field, density, temperature, and speed, as well as increased energy flux. After the sheath, a smooth magnetic field rotation was observed from 22:47 UT on the same day (shaded in yellow). At the same time, solar wind density, temperature, and energy flux decreased, and solar wind speed indicates a declining profile. Therefore, this region corresponds to MC. The FD exhibited a two-step decrease corresponding to these ICME regions, with the rate of decline being highest within the MC (32.4\%).

Additionally, a temporary increase in solar wind density was observed starting at 14:00 UT on March 11, before the ICME’s arrival (indicated in skyblue in Figure \ref{fig:SolO}). A temporal increase followed this in solar wind velocity. These features are typical of a Corotating Interaction Region \citep[CIR:][]{richardson2018solar}, which is an interaction region between fast and slow solar wind \citep[see Figure 1 in][for a clear schematic illustration of general parameter changes in CIR]{kataoka2006flux}. 
Before the ICME reaches the probe, cosmic ray counts often increase gradually before the FD begins, forming a “hill” shape due to the reflection of particles caused by an incoming ICME's shock \citep[e.g.,][]{cane2000coronal, witasse2017interplanetary}. A temporal increase is also observed in the HET data, but it decreases following the arrival of the CIR component. This suggests that the CIR had a shielding effect at Solar Orbiter.

After the MC trailing edge, starting approximately at 07:00 UT on 12 March, the magnetic field became significantly disturbed again, and a sudden increase in electron density to nearly 1000 cm$^{-3}$ occurred (proton density measurements are not available for this interval). This was followed by bipolar variations in some magnetic field components and a subsequent decrease in electron density. These characteristics are typical of the sheath and magnetic ejecta, respectively, suggesting that another ICME had passed through. This ICME, however, has a considerably weaker magnetic field and less clear structure than the previous one and therefore has no significant effect on cosmic ray counts. Thus, we will not discuss it further in this study, but Lavraud et al. (in prep.) provide a detailed analysis of this ICME.

\subsection{Near-Earth spacecraft}\label{sec:wind}
Figure \ref{fig:Earth} presents the observations from spacecraft around Earth as well as from ground bases, which were located 0.99 au from the Sun and were nearly radially aligned with Solar Orbiter (with a longitudinal difference of 10.1 degrees). The comparison of Figures \ref{fig:SolO} and \ref{fig:Earth} shows that overall, the characteristics during the investigated period are similar between Solar Orbiter and Wind. Wind detected the arrival of a sheath at 10:04 UT and the leading edge of the  MC at 22:42 UT on March 13. The CIR peak observed ahead of the ICME in Solar Orbiter data does not appear in Wind observations. However, Wind's solar wind plasma parameters (bulk speed, density, and temperature) show two distinct peaks within the sheath (the purple region in Figure \ref{fig:Earth}). In contrast, Solar Orbiter observed only one peak in the sheath (the purple region in Figure \ref{fig:SolO}).

In ground-based observations, both the Dst and SYM/H indices show a clear storm sudden commencement (SSC) followed by a moderate magnetic storm in terms of its Dst minimum of -85.0 nT (SYM-H momentarily passes the intense storm level, reaching the minimum of -114 nT). The FD observed by the South Pole Neutron Monitor begins almost simultaneously with the onset of disturbances in interplanetary space and on the ground surface, supporting the interpretation that the ICME detected by LRO and Wind corresponds to the same event observed from the Earth. 

During the ICME passage period in Wind, an 18\% decrease was observed in LRO/CRaTER and a 7\% decrease in the South Pole Neutron Monitor. The differences in depth are attributed to differences in the lower energy thresholds (see Section \ref{sec:earth_method}; LRO: 12.7 MeV, Neutron Monitor: 450 MeV). Comparing the FDs observed by LRO/CRaTER and the South Pole Neutron Monitor, both show decreases almost exclusively during the sheath interval and begin to recover during the magnetic cloud phase, with little to no contribution to the decrease from the MC itself. This behavior contrasts with the Solar Orbiter observations, where the main decline occurred within the magnetic cloud. Since the CIR does not precede the ICME, a gradual increase is observed in LRO before the arrival of the ICME, forming a typical “hill” different from that observed by Solar Orbiter.

\subsection{BepiColombo} 
Figure \ref{fig:Bepi-CME} presents the observations from BepiColombo, which was located 0.43 au from the Sun and was offset by 49 degrees in longitude from Solar Orbiter, which was at a nearly equal distance of 0.44 au.
A combined structure consisting of the sheath and the ME arrived at 11:00 UT on March 11. Corresponding to a brief magnetic disturbance, the ion counts measured by the MIA exhibited a slight enhancement in the keV/q range, accompanied by increases in the electron flux, density, and temperature measured by the MEA. These variations are interpreted as signatures of the sheath region. Subsequently, magnetic field observations reveal the presence of an ME structure. The FD exhibited a decrease of approximately 13\%. The second plasma structure arrived at 21:30 UT on the same day (the dashed line in Figure \ref{fig:Bepi-CME}), accompanied by clear enhancements in both electron flux and magnetic field strength. Correspondingly, increases in ion counts were detected in the range of several hundred eV/q to keV/q. However, the SPM count remained unchanged throughout this period. We interpret that the blue and red shaded combined region is basically a single ICME (Figure \ref{fig:Bepi-CME}, the 1st and Mixed). The shock propagating in the middle of it (the dashed line) is due to a sideways interaction with the ICME observed at Solar Orbiter that we have been focusing on, and it is the 2nd ICME for BepiColombo (Figure \ref{fig:Bepi-CME}, the red shaded region). The basis for this interpretation and further discussions is shown in Section \ref{sec:lon}. The arrival of this shock caused the magnetic field to rise rapidly, reaching its peak during this period, but no effect on FD recovery was observed.

Subsequently, at 06:30 UT on 12 March, another magnetic field disturbance was observed, followed by a subtle rotation in the magnetic field rotaion (the yellow shaded 3rd ICME in Figure \ref{fig:Bepi-CME}). This interval featured distinct peaks in both ion counts and electron flux, indicating that BepiColombo may have observed the sheath and magnetic cloud region. This 3rd ICME had a considerably lower magnetic field strength than the previous event, so the SPM count exhibited no significant variations and appeared to be in a recovery phase again. Further discussion of this ICME is provided in Lavraud et al. (in prep.).

From these observations, BepiColombo detected several plasma structures, but only the first one triggered an FD, while the others had little impact.

\section{Discussion} \label{sec:dis}
\subsection{The radial comparison (Solar Orbiter and near Earth spacecraft)} \label{sec:rad}
\subsubsection{The radial decrease in FDs' step number}
\begin{figure}[htbp]
    \centering
    \includegraphics[width=0.95\linewidth]{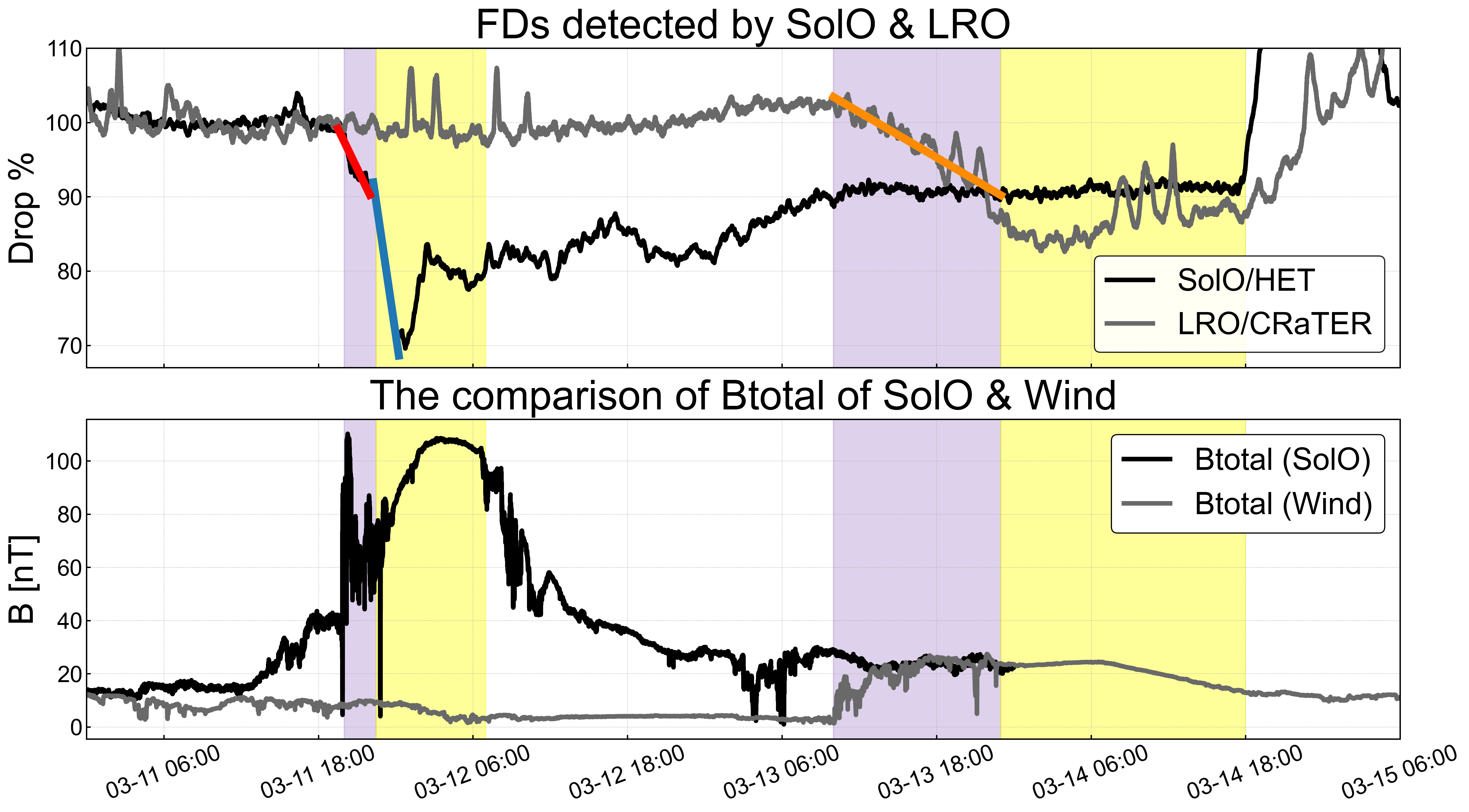}
    \caption{Top: Comparison of FDs observed in nearly straight lines in the heliocentric direction by Solar Orbiter (0.44 au) and LRO (1.0 au). The background is normalized to 100\%, and 16-minute moving averages are taken for appropriate comparison. For Solar Orbiter, the slope of the FD is approximated by linear fitting for each step and indicated in the figure using red and blue lines. For LRO, the FD slope is similarly approximated and shown in orange. Bottom: Comparison of $B_{total}$ from Solar Orbiter and Wind.}
    \label{fig:SolO-Earth}
\end{figure}

Figure \ref{fig:SolO-Earth} shows two FDs observed nearly radially by Solar Orbiter (0.44 au) and LRO (1 au), normalized by their respective pre-ICME background count levels. Reflecting the decay of the ICME during its radial propagation, the FDs' depth decreases from  32.4\% at Solar Orbiter to 18\% at LRO. In addition, consistent with the ICME’s expansion \citep[e.g.,][]{freiherr2020comparing, davies2023characterizing}, the slope of the decrease becomes more gradual. The values labeled as “Slope of the Decrease” in Table \ref{tab:FD_data} represent the rate of decrease (\%/h) determined by linear fitting and indicated in Figure \ref{fig:SolO-Earth} schematically.

One notable aspect of this comparison is that both spacecraft observed nearly the same central portion of a single ICME in succession, capturing both the sheath and the MC, as confirmed by plasma and magnetic field data. However, while a Two-step FD was observed at Solar Orbiter, only a One-step FD was observed at LRO, despite the fact that the ICME at Wind showcased a very clear sheath-MC structure. This contradicts the classical two-step FD model, which attributes the two-step structure to the ICME’s internal components.

In this event, Solar Orbiter passed through the sheath between March 11 19:52-22:47 UT, with a duration of 115 minutes. In contrast, Wind (at LRO) passed through the sheath from March 13 10:04-22:42 UT, with a duration of 758 minutes. The magnetic field magnitude decreased significantly from Solar Orbiter to Wind, from 110 nT to 25 nT, i.e. by 85 nT. \cite{janvier2021two} noted that in ICMEs with well-developed sheaths, substantial screening occurs in the sheath region, reducing the likelihood of further modulation in the following MC (or ME). Based on these observations, we interpret the FD profile at LRO as follows: The prolonged sheath passage at Wind led to a sufficient reduction in count rate during the first step, and the significantly weakened magnetic field in the MC diminished its shielding capacity, resulting in the absence of a significant further FD step and a two-step profile. Therefore, the previously reported trend of decreasing FD step count with heliocentric distance \citep[e.g.,][]{witasse2017interplanetary, winslow2018opening} may be explained by ICME decay and sheath expansion during the propagation. This study is based on a single event analysis; future work will aim to examine similar cases to statistically validate this interpretation.

\begin{table}[h]
 \caption{Characteristics of the FDs at Solar Orbiter, the Earth (LRO and South Pole Neutron Monitor: SoPo), and BepiColombo}
 \centering
 \begin{tabular}{l c c c c c}
 \hline
    & Energy ranges of& Magnitude of &  Gradients of   \\
    & instruments [MeV] & decreases [\%] & decreases [\%/h] \\
 \hline
   Solar Orbiter & $>$16  & 32.4  & grad.1: -3.59, grad.2: -10.2  \\
   Earth (LRO) & $>$12.7  & 18.0  &  All: -1.26, grad.1: -0.83, grad.2: -1.49\\
   Earth (SoPo) & $>$450  & 7.0  &  grad.1: -0.02, grad.2: -0.71\\
   BepiColombo & $>$20  & 13.0  & -3.04 \\
 \hline
 \multicolumn{2}{l}{$^{*}$An ICME interaction occured at BepiColombo (see Section \ref{sec:lon}).} \\
 \multicolumn{4}{l}{$^{*}$All: The gradient assuming one step (Figure \ref{fig:SolO-Earth}); grad. 1 and 2: Assuming two-step (Figures \ref{fig:SolO} and \ref{fig:Earth_FD_two})} \\
 \label{tab:FD_data}
 \end{tabular}
 \end{table}

\subsubsection{Internal structural changes in the sheath and their effects on FDs, as inferred from comparisons in the radial direction}
\cite{laker2024using} focused on the $B_z$ component (They use GSE coordinates) of the magnetic field between Solar Orbiter and Wind  for the same event analyzed here. Figure 4 in \cite{laker2024using} presents Solar Orbiter/MAG, Wind/MFI, Dst, and SYM/H data. Three dip structures in the \( B_z \) component, indicated by arrows in the magnetic field, Dst and SYM/H plot, are observed at both spacecraft. While only the first dip is located ahead of the ICME at Solar Orbiter, all three dips are found within the ICME at Wind. \cite{laker2024using} suggests that a CIR or a small ICME was located ahead of the main ICME, leading to interaction and subsequent incorporation into the sheath. As discussed in Section \ref{sec:solo}, considering that the peak in solar wind density precedes the velocity peak \citep[e.g.,][]{kataoka2006flux, miyoshi2008flux}, we conclude that the front structure is a CIR.

\begin{figure}[h]
    \centering
    \includegraphics[width=0.8\linewidth]{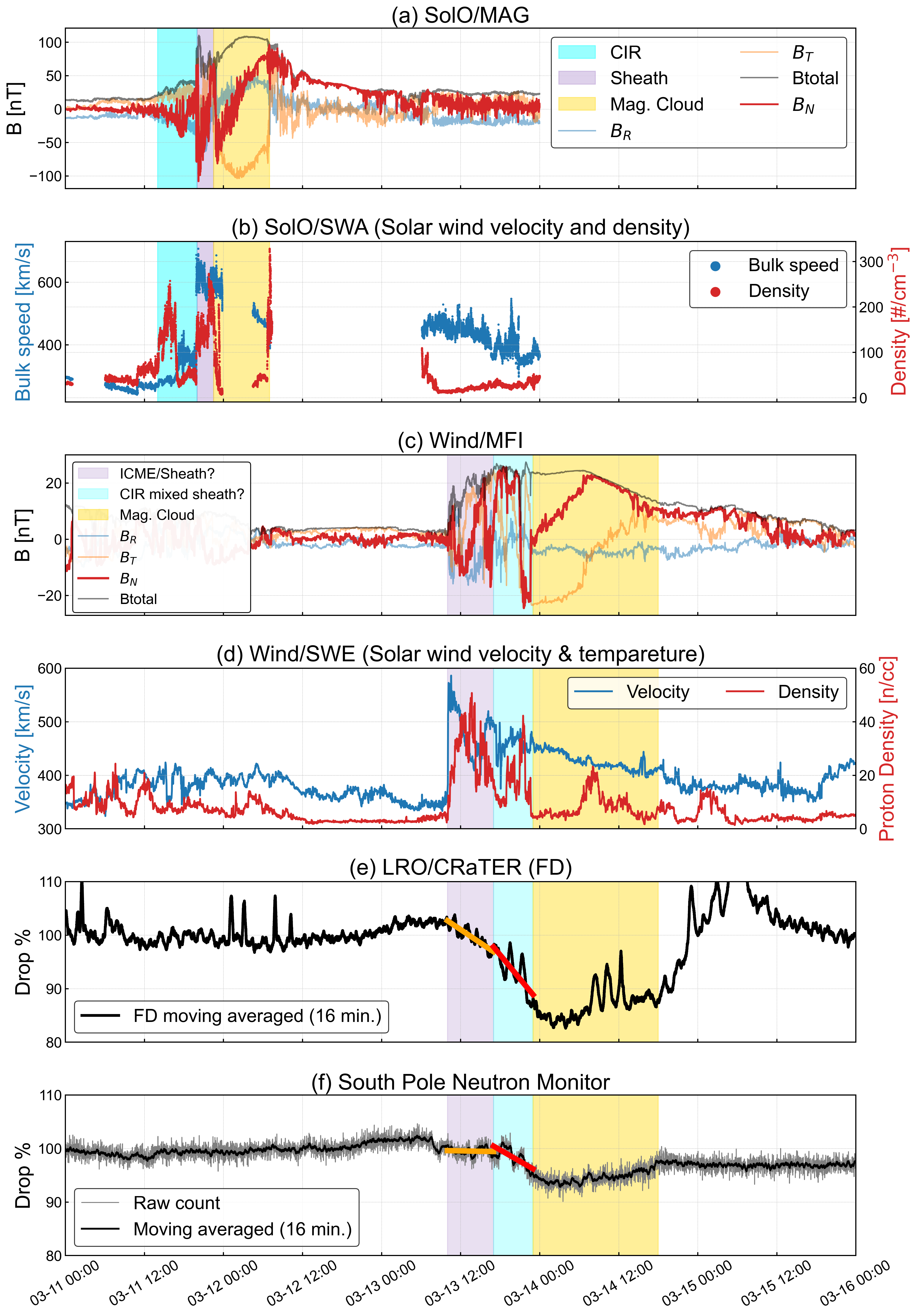}
    \caption{(a): Solar Orbiter's magnetic field data (RTN coordinate system), (b): Solar Orbiter's solar wind data (velocity and density), (c): Wind's magnetic field data (RTN coordinate system), (d): Wind's solar wind data (velocity and density), (e): LRO/CRaTER's FD. The orange and red lines are linear fits, (f): South Pole Neutron Monitor's FD. The orange and red lines are linear fits.}
    \label{fig:Earth_FD_two}
\end{figure}

We focus on the dip in the \( B_N \) component in the RTN coordinate system. At Solar Orbiter, following a sharp increase in the \(-N\) direction within the sheath region, \( B_N \) gradually transitions toward the \( +N \) direction as the MC progresses (Figure \ref{fig:Earth_FD_two}a). In contrast, at Wind, the transition appears as a brief, sudden jump in the \(-N\) direction at the sheath-magnetic cloud boundary. This is immediately followed by a shift to the \( +N \) direction within the MC, where the \(-N \) component is almost entirely absent (Figure \ref{fig:Earth_FD_two}c). These facts strongly suggest that significant interactions had occurred before the MC reached Wind. As stated in Sections \ref{sec:solo} and \ref{sec:wind}, the one peak of solar wind parameters exists both in CIR and ICME/Sheath at Solar Orbiter, while two peaks were observed in ICME/Sheath at Wind. Identifying which of the two solar wind parameter peaks in the sheath at Wind originates from the original ICME or the CIR remains challenging. However, the first peak exhibits a steep profile characteristic of an ICME. In contrast, the second peak has a more gradual increase and a lower absolute value (Figure \ref{fig:Earth_FD_two}d). Therefore, we infer that the first peak originates from the ICME and the second from the CIR.

The coexistence of CIR- and ICME-driven components within the sheath also influences the FD. Since the FD at LRO decreases only within the sheath, it is classified as a one-step FD in the traditional classification. However, as shown in Figure \ref{fig:Earth_FD_two}e, this FD exhibits a slight change in slope between 08:00 and 20:00 UT on March 13 and the period after (the regions marked in purple and blue in Figures \ref{fig:Earth_FD_two}c,d,e,f), corresponding to the peaks in the solar wind parameters. The difference in FD slopes is illustrated using linear fitting. Although it is possible to represent the entire FD with a single slope, as shown in Figure \ref{fig:SolO-Earth}, we use two separate slopes here to highlight the stepwise structure (Figure \ref{fig:Earth_FD_two}e). This indicates that variations in FD gradients arise not only from ICME internal structures but also from differences in the CIR and ICME components. This interpretation is further supported by the FD observed by the ground-based neutron monitor. As shown in Figure \ref{fig:Earth_FD_two}f, the neutron monitor FD also exhibits two slopes within the sheath, in contrast to the smoother profile at LRO. \cite{janvier2021two} noted that particle shielding is energy-dependent: higher-energy particles are more difficult to shield. Since the neutron monitor detects cosmic rays at energies an order of magnitude higher than those measured by LRO/CRaTER (see Section \ref{sec:earth_method}), these particles are correspondingly less affected by shielding. Specifically, during the first peak in the solar wind parameters (the purple region in Figures \ref{fig:Earth_FD_two}c,d,e,f), the shielding was insufficient to cause a significant decrease, resulting in only a moderate drop compared with LRO, and this shock suppressed the preceding ``hill" structure (see Section \ref{sec:insitu}). In contrast, when the second peak arrived (the blue region in Figures \ref{fig:Earth_FD_two}c,d,e,f), a more apparent decrease was observed, leading to the multi-step-like structure.

In summary, the FD gradient profile can indicate not only the sheath-MC structure within ICMEs, but also the interaction between ICMEs and ambient plasma. This study also revealed the effectiveness of comparing data measured at different energy ranges for the same FD in such discussions.

\subsection{The Longitudinal comparison (Solar Orbiter and BepiColombo)} \label{sec:lon}
\begin{figure}
    \centering
    \includegraphics[width=0.65\linewidth]{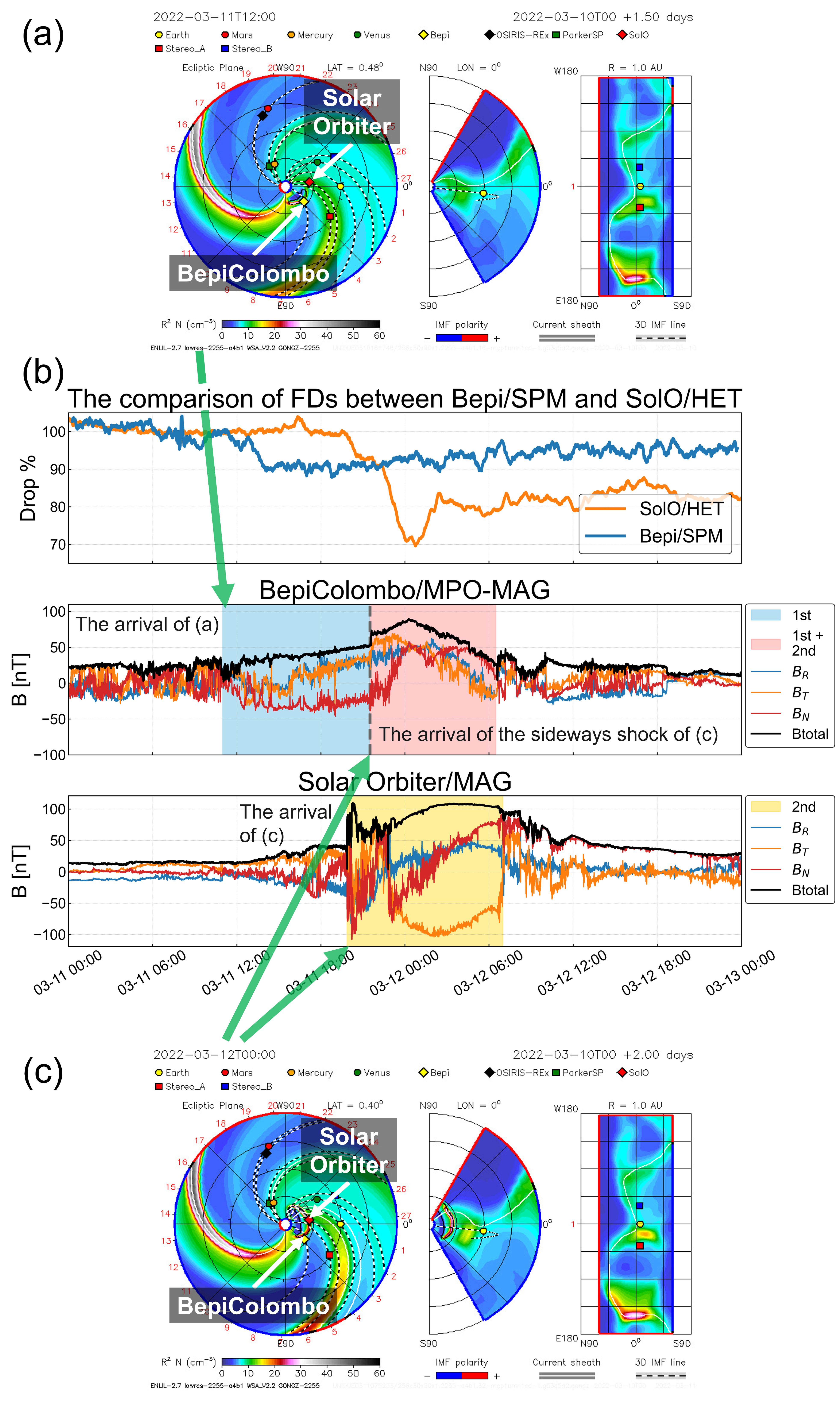}
    \caption{Propagation simulation using the WSA-Enlil + Cone model and in situ observational data for the two ICMEs that occurred in March 2022. (a): First ICME passing BepiColombo (\url{https://kauai.ccmc.gsfc.nasa.gov/DONKI/view/WSA-ENLIL/19412/1}). (b): FDs and magnetic field observed by Solar Orbiter and BepiColombo. The background is normalized to 100\%, and 16-minute moving averages are taken for appropriate comparison. (c): Second ICME observed in coordinated measurements by Solar Orbiter and BepiColombo (the main ICME addressed in this study: \url{https://kauai.ccmc.gsfc.nasa.gov/DONKI/view/WSA-ENLIL/19410/1}).}
    \label{fig:20220310_CME}
\end{figure}

We report on the multi-point observations of an ICME by Solar Orbiter (0.44 au) and BepiColombo (0.43 au). Although the two spacecraft were located at nearly the same heliocentric distance, they were separated by 49° in longitude (Figure \ref{fig:20220310_CME}c). As discussed in Section 4, the ICME was observed at Solar Orbiter at 19:52 UT on 11 March 2022 (Figure \ref{fig:20220310_CME}b, the yellow shaded region) and at BepiColombo at 11:00 UT on the same day (Figure \ref{fig:20220310_CME}b, the combined blue and red shaded region). Lavraud et al. (in prep.) performed a simple Lundquist flux rope model fitting to the magnetic field data from both spacecraft to evaluate the chirality, indicating a left-handed configuration at Solar Orbiter and a right-handed configuration at BepiColombo. The opposite handednesses suggest that the two events were distinct ICMEs. Furthermore, they propose that the shock within the ICME observed by BepiColombo was in fact generated by the interaction with the same ICME detected by Solar Orbiter (see the dashed line in Figure \ref{fig:20220310_CME}b at 21:30 UT on 11 March 2022 in the present study). \cite{koya2024assessment} also suggests that the structure observed by BepiColombo corresponds to the flank of the ICME seen at Solar Orbiter, and the Enlil simulation results also support this (Figures \ref{fig:20220312_cme}, \ref{fig:20220310_CME}c).

Interestingly, despite both spacecraft observing similar peak magnetic field strengths—110 nT at Solar Orbiter and 89.6 nT during the arrival of the second ICME at BepiColombo—the magnitude of the FD was significantly different: 32.4\% at Solar Orbiter, but only 13\% at BepiColombo, with no additional drop associated with the second shock. Therefore, we can interpret that the shielding effect of the second ICME that passed through BepiColombo was small, at less than 13\%.

We interpret this as follows: Solar Orbiter passed through the ICME’s central region (Figure \ref{fig:20220310_CME}c), where the shielding effect on GCR was likely more isotropic, resulting in a larger FD (32.4\%). BepiColombo passed through the central region of the first ICME (Figure \ref{fig:20220310_CME}a), resulting in a 13\% decrease in cosmic ray intensity. Subsequently, it observed the following ICME's shock simultaneously with Solar Orbiter (Figure \ref{fig:20220310_CME}c), but from a different longitudinal position. Since BepiColombo only encountered the edge of the same ICME that passed through Solar Orbiter (the dashed line in Figure \ref{fig:20220310_CME}b), unlike Solar Orbiter, which passed through the core of the same ICME, the shielding effect was anisotropic and weaker. As a result, the cosmic ray decrease caused by the shock (the dashed line in Figure \ref{fig:20220310_CME}b) was not substantial enough to exceed the 13\% decrease caused by the first ICME, and no noticeable change was observed. This interpretation provides observational support for a longitudinal dependence of the GCR shielding effect within a single ICME, as inferred from the comparison between BepiColombo and Solar Orbiter.

\section{Summary \& Conclusions} \label{sec:sum}
Forbush Decreases (FDs) are phenomena in which galactic cosmic rays are temporarily modulated due to solar ejecta such as interplanetary coronal mass ejections (ICMEs). While FDs can be detected using simple particle detectors, characteristics such as their depth, profile, and onset time provide valuable clues for tracing the propagation of ICMEs. Therefore, improving our understanding of the relationship between ICMEs and FDs in interplanetary space is of great significance in space weather.

In this study, we investigated changes in FDs associated with the large-scale evolution of an ICME by comparing multi-point ICME-FD observations obtained in March 2022 by Solar Orbiter, BepiColombo, and near the Earth. Firstly, along the near-radial alignment from the Sun to Earth via Solar Orbiter, we observed progressive changes in the FD characteristics—namely, the decrease in depth, gradient, and number of steps—which reflect the expansion and weakening of ICME substructures during propagation. Furthermore, comparison of FD captured by LRO and ground-based neutron monitors revealed that structural changes caused by interactions between ICME and ambient plasma also affect the gradient of FD. We also conducted a longitudinal comparison between Solar Orbiter and BepiColombo, which were located at nearly the same heliocentric distance but separated by 49 deg. in longitude. However, a direct comparison was hampered by an interaction between two ICMEs observed at BepiColombo. Despite this, an intriguing result was obtained: the second ICME caused a strong magnetic field enhancement at BepiColombo, yet no significant change was observed in the FD. This suggests a longitudinal difference in the shielding effect of the ICME, offering important observational insight.

In statistical analyses of FDs focusing on galactic cosmic ray data acquired by spacecraft in previous studies, the relationship between ICME evolution in the heliocentric radial and longitudinal directions and FD was unclear; however, our results suggest the existence of such relationships. Thus, we think that analysis of galactic cosmic ray variations alone is insufficient for tracking the evolution of ICMEs and FDs in the heliocentric radial and azimuthal directions through multi-point comparisons, and that careful attention must be paid to the correspondence between passing ICME structures and FDs. Specifically, it is crucial to verify the positional relationship of probes on the ICME model and to correlate passing ICME structures and FDs by referencing magnetic field and plasma data.

Observational examples of FDs in the inner heliosphere remain limited, and our understanding of how FDs are initiated and characterized by plasma shortly after their solar ejection is still insufficient. For this reason, datasets of FD observations in the inner heliosphere by missions such as BepiColombo, Solar Orbiter, and Parker Solar Probe \citep{fox2016solar} are increasingly being anticipated in several recent studies \citep[e.g.,][]{davies2023characterizing, belov2023helios}. In this study, we developed techniques for selecting comparable particle detection channels across different spacecraft and established a practical framework for multi-point FD comparisons through case studies. In particular, this study evaluated the characteristics of BepiColombo/SPM and presented a comparison method with other spacecraft for the first time. This study is therefore an important milestone for future FD analysis using BepiColombo. Based on these methods, we aim to analyze additional events and build a statistical foundation for FD research in the inner heliosphere.
\vskip\baselineskip
This work was carried out by the joint research program of the Institute for Space-Earth Environmental Research (ISEE), Nagoya University. G.K. acknowledges the support by JST SPRING, Grant Number JPMJSP2108, and would like to express his gratitude to Dr. Rami Vainio and other members of the BepiColombo/BERM-SPM-SIXS (BESSI) team for their invaluable advice on this research. B.S.-C. acknowledges support through STFC Ernest Rutherford Fellowship ST/V004115/1, and STFC grant ST/Y000439/1. L.R.-G. acknowledges support through the European Space Agency (ESA) research fellowship program. D.H. was supported by the German Ministerium für Wirtschaft und Klimaschutz and the German Zentrum fürLuft‐ und Raumfahrt under contract 50QW2202. French authors acknowledge the French space agency CNES for its support of the Solar Orbiter and BepiColombo missions.
\vskip\baselineskip
We gratefully acknowledge the instrument teams responsible for acquiring the data used in this study and the archive system teams that made the data publicly available. The sources of the data are provided below. 
BepiColombo/SPM data during the period used in this study are archived in the Mio Science Center operated by ISAS/JAXA and ISEE/Nagoya University (\url{https://miosc.isee.nagoya-u.ac.jp/}). BepiColombo/MPO-MAG data is available in ESA's Planetary Science Archive (PSA: \url{https://psa.esa.int/psa/#/pages/home}). Solar Orbiter mission is a project of international cooperation between ESA and NASA, and its data are archived in the Solar Orbiter Archive (SOAR: \url{https://soar.esac.esa.int/soar/}) at ESA's European Space Astronomy Centre (ESAC), Madrid, Spain. We extend our gratitude to the Solar Orbiter/EPD, MAG, and SWA teams for generating and maintaining their data. LRO/CRaTER's data was from NASA/Planetary Data System archive (PDS: \url{https://pds-ppi.igpp.ucla.edu/search/default.jsp#}). Wind/MFI and SWE data were acquired from NASA/Coordinated Data Analysis Web (CDAWeb: \url{https://spdf.gsfc.nasa.gov/pub/data/wind/mfi/}). Global neutron monitor observations are available from the Neutron Monitor Database (NMDB: \url{https://www.nmdb.eu/}).  We acknowledge the World Data Center for Geomagnetism, Kyoto, for providing the Dst and SYM-H indices used in this study. (\url{http://wdc.kugi.kyoto-u.ac.jp/wdc/Sec3.html}).
\vskip\baselineskip
We utilized several Python-based tools in this study, acknowledging the developers. Specifically, ``Sunpy \citep{sunpy_community2020}" was used to depict the image of the Sun and GOES X-ray flux. ``pyIGRF (\url{https://github.com/ciaranbe/pyIGRF})" was employed to evaluate the magnetic field strength around the Earth. PlasmaPy version 2024.10.0, a community-developed, open-source Python package for plasma research and education, was used to calculate lamor radii of galactic cosmic rays around the Earth \citep{plasmapy_community_2024_14010450}. ``Solar-Mach \citep{gieseler2023solar}" was used to calculate the coordinates of the spacecraft. ``SpiceyPy \citep{annex2020spiceypy}" was used to calculate LRO's orbit.

\bibliography{sample631}{}

\begin{thebibliography}{}
\expandafter\ifx\csname natexlab\endcsname\relax\def\natexlab#1{#1}\fi
\providecommand{\url}[1]{\href{#1}{#1}}
\providecommand{\dodoi}[1]{doi:~\href{http://doi.org/#1}{\nolinkurl{#1}}}
\providecommand{\doeprint}[1]{\href{http://ascl.net/#1}{\nolinkurl{http://ascl.net/#1}}}
\providecommand{\doarXiv}[1]{\href{https://arxiv.org/abs/#1}{\nolinkurl{https://arxiv.org/abs/#1}}}

\bibitem[{Allison {et~al.}(2016)Allison, Amako, Apostolakis, Arce, Asai, Aso, Bagli, Bagulya, Banerjee, Barrand, {et~al.}}]{allison2016recent}
Allison, J., Amako, K., Apostolakis, J., {et~al.} 2016, Nuclear instruments and methods in physics research section A: Accelerators, Spectrometers, Detectors and Associated Equipment, 835, 186, \dodoi{https://doi.org/10.1016/j.nima.2016.06.125.}

\bibitem[{Andrews {et~al.}(2007)Andrews, Zurbuchen, Mauk, Malcom, Fisk, Gloeckler, Ho, Kelley, Koehn, LeFevere, {et~al.}}]{andrews2007energetic}
Andrews, G.~B., Zurbuchen, T.~H., Mauk, B.~H., {et~al.} 2007, Space science reviews, 131, 523, \dodoi{https://doi.org/10.1007/s11214-007-9272-5}

\bibitem[{Annex {et~al.}(2020)Annex, Pearson, Seignovert, Carcich, Eichhorn, Mapel, Von~Forstner, McAuliffe, Del~Rio, Berry, {et~al.}}]{annex2020spiceypy}
Annex, A.~M., Pearson, B., Seignovert, B., {et~al.} 2020, Journal of Open Source Software, 5, 2050

\bibitem[{Arge {et~al.}(2004)Arge, Luhmann, Odstrcil, Schrijver, \& Li}]{arge2004stream}
Arge, C., Luhmann, J., Odstrcil, D., Schrijver, C., \& Li, Y. 2004, Journal of Atmospheric and Solar-Terrestrial Physics, 66, 1295, \dodoi{https://doi.org/10.1016/j.jastp.2004.03.018}

\bibitem[{Arge \& Pizzo(2000)}]{arge2000improvement}
Arge, C., \& Pizzo, V. 2000, Journal of Geophysical Research: Space Physics, 105, 10465, \dodoi{https://doi.org/10.1029/1999JA000262}

\bibitem[{Baumjohann {et~al.}(2020)Baumjohann, Matsuoka, Narita, Magnes, Heyner, Glassmeier, Nakamura, Fischer, Plaschke, Volwerk, {et~al.}}]{baumjohann2020bepicolombo}
Baumjohann, W., Matsuoka, A., Narita, Y., {et~al.} 2020, Space Science Reviews, 216, 125, \dodoi{https://doi.org/10.1007/s11214-020-00754-y}

\bibitem[{Belov {et~al.}(2023)Belov, Shlyk, Abunina, Abunin, Papaioannou, Richardson, \& Lario}]{belov2023helios}
Belov, A., Shlyk, N., Abunina, M., {et~al.} 2023, Monthly Notices of the Royal Astronomical Society, 521, 4652, \dodoi{10.1093/mnras/stad732}

\bibitem[{Benkhoff {et~al.}(2021)Benkhoff, Murakami, Baumjohann, Besse, Bunce, Casale, Cremosese, Glassmeier, Hayakawa, Heyner, {et~al.}}]{benkhoff2021bepicolombo}
Benkhoff, J., Murakami, G., Baumjohann, W., {et~al.} 2021, Space science reviews, 217, 90, \dodoi{https://doi.org/10.1007/s11214-021-00861-4}

\bibitem[{Burlaga {et~al.}(1981)Burlaga, Sittler, Mariani, \& Schwenn}]{burlaga1981magnetic}
Burlaga, L., Sittler, E., Mariani, F., \& Schwenn, a.~R. 1981, Journal of Geophysical Research: Space Physics, 86, 6673

\bibitem[{Cane(2000)}]{cane2000coronal}
Cane, H.~V. 2000, Space Science Reviews, 93, 55, \dodoi{https://doi.org/10.1023/A:1026532125747}

\bibitem[{Cane {et~al.}(1994)Cane, Richardson, Von~Rosenvinge, \& Wibberenz}]{cane1994cosmic}
Cane, H.~V., Richardson, I., Von~Rosenvinge, T., \& Wibberenz, G. 1994, Journal of Geophysical Research: Space Physics, 99, 21429, \dodoi{https://doi.org/10.1029/94JA01529}

\bibitem[{CE(1997)}]{ce1997international}
CE, B. 1997, Journal of geomagnetism and geoelectricity, 49, 123, \dodoi{https://doi.org/10.5636/jgg.49.123}

\bibitem[{Clem \& Dorman(2000)}]{clem2000neutron}
Clem, J.~M., \& Dorman, L.~I. 2000, in Cosmic Rays and Earth: Proceedings of an ISSI Workshop, 21--26 March 1999, Bern, Switzerland, Springer, 335--359, \dodoi{https://ui.adsabs.harvard.edu/link_gateway/2000SSRv...93..335C/doi:10.1023/A:1026508915269}

\bibitem[{Davies {et~al.}(2023)Davies, Winslow, \& Lawrence}]{davies2023characterizing}
Davies, E.~E., Winslow, R.~M., \& Lawrence, D.~J. 2023, The Astrophysical Journal, 943, 83, \dodoi{10.3847/1538-4357/aab098}

\bibitem[{Davies {et~al.}(2021)Davies, M{\"o}stl, Owens, Weiss, Amerstorfer, Hinterreiter, Bauer, Bailey, Reiss, Forsyth, {et~al.}}]{davies2021situ}
Davies, E.~E., M{\"o}stl, C., Owens, M., {et~al.} 2021, Astronomy \& Astrophysics, 656, A2, \dodoi{https://doi.org/10.1051/0004-6361/202040113}

\bibitem[{Forbush(1937)}]{forbush1937effects}
Forbush, S.~E. 1937, Physical Review, 51, 1108, \dodoi{https://doi.org/10.1103/PhysRev.51.1108.3}

\bibitem[{Fox {et~al.}(2016)Fox, Velli, Bale, Decker, Driesman, Howard, Kasper, Kinnison, Kusterer, Lario, {et~al.}}]{fox2016solar}
Fox, N., Velli, M., Bale, S., {et~al.} 2016, Space Science Reviews, 204, 7, \dodoi{https://doi.org/10.1007/s11214-015-0211-6}

\bibitem[{Fraenz {et~al.}(2024)Fraenz, Rojo, Cornet, Hadid, Saito, Andr{\'e}, Varsani, Schmid, Krueger, Krupp, {et~al.}}]{fraenz2024spacecraft}
Fraenz, M., Rojo, M., Cornet, T., {et~al.} 2024, Journal of Geophysical Research: Space Physics, 129, e2023JA032044, \dodoi{https://doi.org/10.1029/2023JA032044}

\bibitem[{Freiherr~von Forstner {et~al.}(2020)Freiherr~von Forstner, Guo, Wimmer-Schweingruber, Dumbovi{\'c}, Janvier, D{\'e}moulin, Veronig, Temmer, Papaioannou, Dasso, {et~al.}}]{freiherr2020comparing}
Freiherr~von Forstner, J.~L., Guo, J., Wimmer-Schweingruber, R.~F., {et~al.} 2020, Journal of Geophysical Research: Space Physics, 125, e2019JA027662, \dodoi{https://doi.org/10.1029/2019JA027662}

\bibitem[{Gieseler {et~al.}(2023)Gieseler, Dresing, Palmroos, Freiherr~von Forstner, Price, Vainio, Kouloumvakos, Rodr{\'\i}guez-Garc{\'\i}a, Trotta, G{\'e}not, {et~al.}}]{gieseler2023solar}
Gieseler, J., Dresing, N., Palmroos, C., {et~al.} 2023, Frontiers in Astronomy and Space Sciences, 9, 1058810, \dodoi{https://doi.org/10.3389/fspas.2022.1058810}

\bibitem[{Harada {et~al.}(2022)Harada, Aizawa, Saito, Andr{\'e}, Persson, Delcourt, Hadid, Fraenz, Yokota, Fedorov, {et~al.}}]{harada2022bepicolombo}
Harada, Y., Aizawa, S., Saito, Y., {et~al.} 2022, Geophysical Research Letters, 49, e2022GL100279, \dodoi{https://doi.org/10.1029/2022GL100279}

\bibitem[{Harada {et~al.}(2024)Harada, Saito, Hadid, Delcourt, Aizawa, Rojo, Andr{\'e}, Persson, Fraenz, Yokota, {et~al.}}]{harada2024deep}
Harada, Y., Saito, Y., Hadid, L.~Z., {et~al.} 2024, Journal of Geophysical Research: Space Physics, 129, e2024JA032751, \dodoi{https://doi.org/10.1029/2024JA032751}

\bibitem[{Heyner {et~al.}(2021)Heyner, Auster, Forna{\c{c}}on, Carr, Richter, Mieth, Kolhey, Exner, Motschmann, Baumjohann, {et~al.}}]{heyner2021bepicolombo}
Heyner, D., Auster, H.-U., Forna{\c{c}}on, K.-H., {et~al.} 2021, Space Science Reviews, 217, 1, \dodoi{https://doi.org/10.1007/s11214-021-00822-x}

\bibitem[{Horbury {et~al.}(2020)Horbury, O’brien, Blazquez, Bendyk, Brown, Hudson, Evans, Oddy, Carr, Beek, {et~al.}}]{horbury2020solar}
Horbury, T., O’brien, H., Blazquez, I.~C., {et~al.} 2020, Astronomy \& Astrophysics, 642, A9, \dodoi{https://doi.org/10.1051/0004-6361/201937257}

\bibitem[{Jackson {et~al.}(2023)Jackson, Tokumaru, Iwai, Bracamontes, Buffington, Fujiki, Murakami, Heyner, Sanchez-Cano, Rojo, {et~al.}}]{jackson2023forecasting}
Jackson, B.~V., Tokumaru, M., Iwai, K., {et~al.} 2023, Solar Physics, 298, 74, \dodoi{https://doi.org/10.1007/s11207-023-02169-8}

\bibitem[{Janvier {et~al.}(2021)Janvier, D{\'e}moulin, Guo, Dasso, Regnault, Topsi-Moutesidou, Gutierrez, \& Perri}]{janvier2021two}
Janvier, M., D{\'e}moulin, P., Guo, J., {et~al.} 2021, The Astrophysical Journal, 922, 216, \dodoi{https://doi.org/10.3847/1538-4357/ac2b9b}

\bibitem[{Jordan {et~al.}(2011)Jordan, Spence, Blake, \& Shaul}]{jordan2011revisiting}
Jordan, A.~P., Spence, H.~E., Blake, J., \& Shaul, D. 2011, Journal of Geophysical Research: Space Physics, 116, \dodoi{https://doi.org/10.1029/2011JA016791}

\bibitem[{Kaiser {et~al.}(2008)Kaiser, Kucera, Davila, St.~Cyr, Guhathakurta, \& Christian}]{kaiser2008stereo}
Kaiser, M.~L., Kucera, T., Davila, J., {et~al.} 2008, Space Science Reviews, 136, 5, \dodoi{https://doi.org/10.1007/s11214-007-9277-0}

\bibitem[{Kataoka \& Miyoshi(2006)}]{kataoka2006flux}
Kataoka, R., \& Miyoshi, Y. 2006, Space Weather, 4, \dodoi{https://doi.org/10.1029/2005SW000211}

\bibitem[{Khotyaintsev {et~al.}(2021)Khotyaintsev, Graham, Vaivads, Steinvall, Edberg, Eriksson, Johansson, Sorriso-Valvo, Maksimovic, Bale, {et~al.}}]{khotyaintsev2021density}
Khotyaintsev, Y.~V., Graham, D.~B., Vaivads, A., {et~al.} 2021, Astronomy \& Astrophysics, 656, A19, \dodoi{https://doi.org/10.1051/0004-6361/202140936}

\bibitem[{Kilpua {et~al.}(2017)Kilpua, Koskinen, \& Pulkkinen}]{kilpua2017coronal}
Kilpua, E., Koskinen, H.~E., \& Pulkkinen, T.~I. 2017, Living Reviews in Solar Physics, 14, 1, \dodoi{https://doi.org/10.1007/s41116-017-0009-6}

\bibitem[{Kinoshita {et~al.}(2025)Kinoshita, Ueno, Murakami, Pinto, Yoshioka, \& Miyoshi}]{kinoshita2025simulation}
Kinoshita, G., Ueno, H., Murakami, G., {et~al.} 2025, Journal of Geophysical Research: Space Physics, 130, e2024JA033147, \dodoi{https://doi.org/10.1029/2024JA033147}

\bibitem[{Koya {et~al.}(2024)Koya, Patsourakos, Georgoulis, \& Nindos}]{koya2024assessment}
Koya, S., Patsourakos, S., Georgoulis, M.~K., \& Nindos, A. 2024, Astronomy \& Astrophysics, 690, A233, \dodoi{https://doi.org/10.1051/0004-6361/202450204}

\bibitem[{Laker {et~al.}(2024)Laker, Horbury, O’Brien, Fauchon-Jones, Angelini, Fargette, Amerstorfer, Bauer, M{\"o}stl, Davies, {et~al.}}]{laker2024using}
Laker, R., Horbury, T., O’Brien, H., {et~al.} 2024, Space Weather, 22, e2023SW003628, \dodoi{https://doi.org/10.1029/2023SW003628}

\bibitem[{Lepping {et~al.}(1995)Lepping, Ac{\~u}na, Burlaga, Farrell, Slavin, Schatten, Mariani, Ness, Neubauer, Whang, {et~al.}}]{lepping1995wind}
Lepping, R., Ac{\~u}na, M., Burlaga, L., {et~al.} 1995, Space Science Reviews, 71, 207, \dodoi{https://doi.org/10.1007/BF00751330}

\bibitem[{Maksimovic {et~al.}(2020)Maksimovic, Bale, Chust, Khotyaintsev, Krasnoselskikh, Kretzschmar, Plettemeier, Rucker, Sou{\v{c}}ek, Steller, {et~al.}}]{maksimovic2020solar}
Maksimovic, M., Bale, S., Chust, T., {et~al.} 2020, Astronomy \& Astrophysics, 642, A12, \dodoi{https://doi.org/10.1051/0004-6361/201936214}

\bibitem[{Millward {et~al.}(2013)Millward, Biesecker, Pizzo, \& De~Koning}]{millward2013operational}
Millward, G., Biesecker, D., Pizzo, V., \& De~Koning, C. 2013, Space Weather, 11, 57, \dodoi{https://doi.org/10.1002/swe.20024}

\bibitem[{Miyoshi \& Kataoka(2005)}]{miyoshi2005ring}
Miyoshi, Y., \& Kataoka, R. 2005, Geophysical research letters, 32, \dodoi{https://doi.org/10.1029/2005GL024590}

\bibitem[{Miyoshi \& Kataoka(2008)}]{miyoshi2008flux}
---. 2008, Journal of Geophysical Research: Space Physics, 113, \dodoi{https://doi.org/10.1029/2007JA012506}

\bibitem[{M{\"u}ller {et~al.}(2020)M{\"u}ller, Cyr, Zouganelis, Gilbert, Marsden, Nieves-Chinchilla, Antonucci, Auchere, Berghmans, Horbury, {et~al.}}]{muller2020solar}
M{\"u}ller, D., Cyr, O.~S., Zouganelis, I., {et~al.} 2020, Astronomy \& Astrophysics, 642, A1, \dodoi{https://doi.org/10.1051/0004-6361/202038467}

\bibitem[{Murakami {et~al.}(2020)Murakami, Hiroyuki, Shoya, Taeko, Yasumasa, Yoshifumi, Ichiro, Masanori, Wolfgang, Ayako, {et~al.}}]{go2020mio}
Murakami, G., Hiroyuki, O., Shoya, M., {et~al.} 2020, Space Science Reviews, 216, \dodoi{https://doi.org/10.1007/s11214-020-00733-3}

\bibitem[{Odstrcil(2003)}]{odstrcil2003modeling}
Odstrcil, D. 2003, Advances in Space Research, 32, 497, \dodoi{https://doi.org/10.1016/S0273-1177(03)00332-6}

\bibitem[{Ogilvie \& Desch(1997)}]{ogilvie1997wind}
Ogilvie, K., \& Desch, M. 1997, Advances in Space Research, 20, 559, \dodoi{https://doi.org/10.1016/S0273-1177(97)00439-0}

\bibitem[{Ogilvie {et~al.}(1995)Ogilvie, Chornay, Fritzenreiter, Hunsaker, Keller, Lobell, Miller, Scudder, Sittler, Torbert, {et~al.}}]{ogilvie1995swe}
Ogilvie, K., Chornay, D., Fritzenreiter, R., {et~al.} 1995, Space Science Reviews, 71, 55, \dodoi{https://doi.org/https://doi.org/10.1007/BF00751326}

\bibitem[{Owen {et~al.}(2020)Owen, Bruno, Livi, Louarn, Al~Janabi, Allegrini, Amoros, Baruah, Barthe, Berthomier, {et~al.}}]{owen2020solar}
Owen, C., Bruno, R., Livi, S., {et~al.} 2020, Astronomy \& Astrophysics, 642, A16, \dodoi{https://doi.org/10.1051/0004-6361/201937259}

\bibitem[{Owens {et~al.}(2017)Owens, Lockwood, \& Barnard}]{owens2017coronal}
Owens, M., Lockwood, M., \& Barnard, L. 2017, Scientific Reports, 7, 4152, \dodoi{https://doi.org/10.1038/s41598-017-04546-3}

\bibitem[{Palmerio {et~al.}(2024)Palmerio, Carcaboso, Khoo, Salman, S{\'a}nchez-Cano, Lynch, Rivera, Pal, Nieves-Chinchilla, Weiss, {et~al.}}]{palmerio2024mesoscale}
Palmerio, E., Carcaboso, F., Khoo, L.~Y., {et~al.} 2024, The Astrophysical Journal, 963, 108, \dodoi{https://doi.org/10.3847/1538-4357/ad1ab4}

\bibitem[{{PlasmaPy Community} {et~al.}(2024){PlasmaPy Community}, Murphy, Everson, Stańczak-Marikin, Heuer, Kozlowski, Johnson, Malhotra, Schaffner, Vincena, Abler, Addison, Ahamed, Alarcon, Antognetti, Arran, Bagherianlemraski, Beckers, Bedmutha, Bedoya-Lopez, Bergeron, Bessi, Britten, Brown, Bryant, Carroll, Cartagena-Sanchez, Chambers, Chattopadhyay, Choubey, Choudhary, Clauss, Colom, Davies, Deal, Decristoforo, Diaz~Riega, Dover, Drozdov, Du, Einhorn, Ervin, Fan, Farid, Fischer, Foo, Fütterer, Gangadharan, Gerow, Gonzalez, Goodall, Gorelli, Gordon-McKeon, Goudeau, Guidoni, Guimiot, Haggerty, Hansen, Haque, Hillairet, Hoang, How, Huang, Humphrey, Isupova, Jeandet, Jones, Kastek, Kent, Klima, Köhn-Seemann, Kulshrestha, Kumar, Kuszaj, Langendorf, Lanteri, Lee, Leonard, Lequette, Lim, Magarde, Martinelli, Masood, McHardy, Modi, Montes, Mumford, Munn, Murphy, Nie, Ortiz, Panda, Pannala, Parashar, Patel, Pavon, Pérez, Pitzer, Polak, Qudsi, Raj, Rajashekar, Rao, Reep, Richardson, Roberts, Rodriguez,
  Rojas~Zelaya, Salcido, Savcheva, Schneck, Shen, Sheng, Sherpa, Silvestri, Simon, Singh, Singh, Sipőcz, Skinner, Skrzypczak, Smirnov, Smith, Sobeske, Spedicato, Stansby, Stinson, Sugiharto, Švancarová, Tavant, Tranquilino, Ulrich, Valle, Varnish, Vo, Wu, Xu, Yip, \& Zhang}]{plasmapy_community_2024_14010450}
{PlasmaPy Community}, Murphy, N.~A., Everson, E.~T., {et~al.} 2024, PlasmaPy, v2024.10.0,  Zenodo, \dodoi{10.5281/zenodo.14010450}

\bibitem[{Richardson(2018)}]{richardson2018solar}
Richardson, I.~G. 2018, Living reviews in solar physics, 15, 1, \dodoi{https://doi.org/10.1007/s41116-017-0011-z}

\bibitem[{Rodr{\'\i}guez-Garc{\'\i}a {et~al.}(2022)Rodr{\'\i}guez-Garc{\'\i}a, Nieves-Chinchilla, G{\'o}mez-Herrero, Zouganelis, Vourlidas, Balmaceda, Dumbovi{\'c}, Jian, Mays, Carcaboso, {et~al.}}]{rodriguez2022evidence}
Rodr{\'\i}guez-Garc{\'\i}a, L., Nieves-Chinchilla, T., G{\'o}mez-Herrero, R., {et~al.} 2022, Astronomy \& astrophysics, 662, A45, \dodoi{https://doi.org/10.1051/0004-6361/202142966}

\bibitem[{Rodr{\'\i}guez-Pacheco {et~al.}(2020)Rodr{\'\i}guez-Pacheco, Wimmer-Schweingruber, Mason, Ho, S{\'a}nchez-Prieto, Prieto, Mart{\'\i}n, Seifert, Andrews, Kulkarni, {et~al.}}]{rodriguez2020energetic}
Rodr{\'\i}guez-Pacheco, J., Wimmer-Schweingruber, R., Mason, G., {et~al.} 2020, Astronomy \& Astrophysics, 642, A7, \dodoi{https://doi.org/10.1051/0004-6361/201935287}

\bibitem[{Rojo {et~al.}(2024)Rojo, Persson, Sauvaud, Aizawa, Nicolaou, Penou, Barthe, Andr{\'e}, Mazelle, Fedorov, {et~al.}}]{rojo2024electron}
Rojo, M., Persson, M., Sauvaud, J.-A., {et~al.} 2024, Astronomy \& Astrophysics, 683, A99, \dodoi{https://doi.org/10.1051/0004-6361/202347843}

\bibitem[{Saito {et~al.}(2021)Saito, Delcourt, Hirahara, Barabash, Andr{\'e}, Takashima, Asamura, Yokota, Wieser, Nishino, {et~al.}}]{saito2021pre}
Saito, Y., Delcourt, D., Hirahara, M., {et~al.} 2021, Space Science Reviews, 217, 1, \dodoi{https://doi.org/10.1007/s11214-021-00839-2}

\bibitem[{Sanchez-Cano {et~al.}(2025)Sanchez-Cano, Hadid, Aizawa, Murakami, Bamba, \& Chiba}]{beatriz2025cruise}
Sanchez-Cano, B., Hadid, L., Aizawa, S., {et~al.} 2025, Earth, Planets and Space, 77, 114, \dodoi{https://doi.org/10.1186/s40623-025-02256-z}

\bibitem[{Simpson(1983)}]{simpson1983elemental}
Simpson, J. 1983, Ann. Rev. Nucl. Part. Sci.;(United States), 33, \dodoi{https://doi.org/10.1146/annurev.ns.33.120183.001543}

\bibitem[{Solomon {et~al.}(2007)Solomon, McNutt, Gold, \& Domingue}]{solomon2007messenger}
Solomon, S.~C., McNutt, R.~L., Gold, R.~E., \& Domingue, D.~L. 2007, Space Science Reviews, 131, 3, \dodoi{https://doi.org/10.1007/s11214-007-9247-6}

\bibitem[{Spence {et~al.}(2010)Spence, Case, Golightly, Heine, Larsen, Blake, Caranza, Crain, George, Lalic, {et~al.}}]{spence2010crater}
Spence, H.~E., Case, A., Golightly, M., {et~al.} 2010, Space science reviews, 150, 243, \dodoi{https://doi.org/10.1007/s11214-009-9584-8}

\bibitem[{Sullivan(1971)}]{sullivan1971geometric}
Sullivan, J. 1971, Nuclear Instruments and methods, 95, 5, \dodoi{https://doi.org/10.1016/0029-554X(71)90033-4}

\bibitem[{{The SunPy Community} {et~al.}(2020){The SunPy Community}, Barnes, Bobra, Christe, Freij, Hayes, Ireland, Mumford, Perez-Suarez, Ryan, Shih, Chanda, Glogowski, Hewett, Hughitt, Hill, Hiware, Inglis, Kirk, Konge, Mason, Maloney, Murray, Panda, Park, Pereira, Reardon, Savage, Sipőcz, Stansby, Jain, Taylor, Yadav, Rajul, \& Dang}]{sunpy_community2020}
{The SunPy Community}, Barnes, W.~T., Bobra, M.~G., {et~al.} 2020, The Astrophysical Journal, 890, 68, \dodoi{10.3847/1538-4357/ab4f7a}

\bibitem[{von Forstner {et~al.}(2021)von Forstner, Dumbovi{\'c}, M{\"o}stl, Guo, Papaioannou, Elftmann, Xu, Terasa, Kollhoff, Wimmer-Schweingruber, {et~al.}}]{von2021radial}
von Forstner, J. L.~F., Dumbovi{\'c}, M., M{\"o}stl, C., {et~al.} 2021, Astronomy \& astrophysics, 656, A1, \dodoi{https://doi.org/10.1051/0004-6361/202039848}

\bibitem[{Vondrak {et~al.}(2010)Vondrak, Keller, Chin, \& Garvin}]{vondrak2010lunar}
Vondrak, R., Keller, J., Chin, G., \& Garvin, J. 2010, Space science reviews, 150, 7, \dodoi{https://doi.org/10.1007/s11214-010-9631-5}

\bibitem[{Wanliss \& Showalter(2006)}]{wanliss2006high}
Wanliss, J.~A., \& Showalter, K.~M. 2006, Journal of Geophysical Research: Space Physics, 111, \dodoi{https://doi.org/10.1029/2005JA011034}

\bibitem[{Wibberenz {et~al.}(1998)Wibberenz, Le~Roux, Potgieter, \& Bieber}]{wibberenz1998transient}
Wibberenz, G., Le~Roux, J., Potgieter, M., \& Bieber, J. 1998, Space science reviews, 83, 309, \dodoi{https://doi.org/10.1023/A:1005083109827}

\bibitem[{Wilson {et~al.}(2020)Wilson, Spence, Schwadron, Case, Looper, Jordan, de~Wet, \& Kasper}]{wilson2020precise}
Wilson, J.~K., Spence, H.~E., Schwadron, N.~A., {et~al.} 2020, Geophysical research letters, 47, e2019GL085522, \dodoi{https://doi.org/10.1029/2019GL085522}

\bibitem[{Winslow {et~al.}(2015)Winslow, Lugaz, Philpott, Schwadron, Farrugia, Anderson, \& Smith}]{winslow2015interplanetary}
Winslow, R.~M., Lugaz, N., Philpott, L.~C., {et~al.} 2015, Journal of Geophysical Research: Space Physics, 120, 6101, \dodoi{https://doi.org/10.1002/2015JA021200}

\bibitem[{Winslow {et~al.}(2018)Winslow, Schwadron, Lugaz, Guo, Joyce, Jordan, Wilson, Spence, Lawrence, Wimmer-Schweingruber, {et~al.}}]{winslow2018opening}
Winslow, R.~M., Schwadron, N.~A., Lugaz, N., {et~al.} 2018, The Astrophysical Journal, 856, 139, \dodoi{10.3847/1538-4357/aab098}

\bibitem[{Witasse {et~al.}(2017)Witasse, S{\'a}nchez-Cano, Mays, Kajdi{\v{c}}, Opgenoorth, Elliott, Richardson, Zouganelis, Zender, Wimmer-Schweingruber, {et~al.}}]{witasse2017interplanetary}
Witasse, O., S{\'a}nchez-Cano, B., Mays, M., {et~al.} 2017, Journal of Geophysical Research: Space Physics, 122, 7865, \dodoi{https://doi.org/10.1002/2017JA023884}

\bibitem[{Xie {et~al.}(2004)Xie, Ofman, \& Lawrence}]{xie2004cone}
Xie, H., Ofman, L., \& Lawrence, G. 2004, Journal of Geophysical Research: Space Physics, 109, \dodoi{https://doi.org/10.1029/2003JA010226}

\bibitem[{Zhuang {et~al.}(2024)Zhuang, Lugaz, Al-Haddad, Scolini, Farrugia, Regnault, Davies, Yu, Winslow, \& Galvin}]{zhuang2024combining}
Zhuang, B., Lugaz, N., Al-Haddad, N., {et~al.} 2024, Astronomy \& Astrophysics, 682, A107, \dodoi{https://doi.org/10.1051/0004-6361/202347561}

\bibitem[{Zurbuchen \& Richardson(2006)}]{kunow2006situ}
Zurbuchen, T., \& Richardson, I. 2006, Coronal mass ejections, 31, \dodoi{https://doi.org/10.1007/s11214-006-9010-4}

\end{thebibliography}
\bibliographystyle{aasjournal}



\end{document}